\documentclass{article}

\usepackage{PRIMEarxiv}
\usepackage{subfig}
\usepackage[utf8]{inputenc} 
\usepackage[T1]{fontenc}    
\usepackage{hyperref}       
\usepackage{url}            
\usepackage{booktabs}       
\usepackage{amsfonts}       
\usepackage{nicefrac}       
\usepackage{microtype}      
\usepackage{lipsum}
\usepackage{fancyhdr}       
\usepackage{graphicx}       
\graphicspath{{media/}}     
\usepackage{amsmath}
\title{DiffusionVel: Multi-Information Integrated Velocity Inversion Using Generative Diffusion Models
}

\author{
  Hao Zhang \\
   China University of Petroleum (East China)\\
  \texttt{haozhang.real@gmail.com} \\
   \And
  Yuanyuan Li \\
   China University of Petroleum (East China)\\
  \texttt{yuanyuanli@upc.edu.cn} \\
     \And
  Jianping Huang \\
   China University of Petroleum (East China)\\
  \texttt{jphuang@upc.edu.cn)} \\
}

\begin{document}
\maketitle

\begin{abstract}
Full waveform inversion (FWI) is capable of reconstructing subsurface properties with high resolution from seismic data. However, conventional FWI faces challenges such as cycle-skipping and high computational costs. Recently, deep learning method has emerged as a promising solution for efficient velocity estimation. We develop DiffusionVel, a data-driven technique based on the state-of-the-art generative diffusion models (GDMs) with integration of multiple information including seismic data, background velocity, geological knowledge, and well logs. We use two separate conditional GDMs, namely the seismic-data GDM and the well-log GDM, and an unconditional GDM, i.e., the geology-oriented GDM, to adapt the generated velocity model to the constraints of seismic data, well logs, and prior geological knowledge, respectively. Besides, the background velocity can be incorporated into the generated velocity model with a low-pass filter. The generation of these GDM are then combined together with a weighted summation in the sampling process. We can flexibly control the constraints from each information by adjusting the weighting factors. We make a comprehensive comparison between the proposed DiffusionVel and three previously-developed methods including conventional FWI, InversionNet, and VelocityGAN by using the OpenFWI datasets and the Hess VTI model example. The test results demonstrate that the proposed DiffusionVel method predicts the velocity model reasonably by integrating multiple information effectively.
\end{abstract}


\section{Introduction}
Seismic full waveform inversion (FWI) has emerged as an advanced technique for estimating subsurface properties with high resolution \cite{tarantola1984inversion}. The subsurface model is updated iteratively by optimizing the data matching between simulated and observed seismic data \cite{virieux2009overview}. FWI is typically an ill-posed inverse problem due to the band-limited seismic data and the limitations of seismic acquisition geometries. In particular, FWI often suffers from the cycle-skipping issue, and thus getting trapped in a local minima, when the simulated data is more than half cycle away from the observed data. A typical implementation of FWI involves forward modeling to predict the seismic data for the updated model, optimization algorithm, gradient computation and regularization techniques to handle the ill-posed inverse problem. The regularization techniques allow us to introduce additional constraints or penalties into the inversion process to guide the inversion towards more reasonable and stable solution. This can include constraints on model smoothness (Tikhonov regularization;\cite{golub1999tikhonov}), sparsity (Total variation regularization;\cite{strong2003edge}), or prior knowledge about the subsurface model \cite{asnaashari2013regularized,zhang2022regularized}. Besides, multiple types of data or measurements, such as well logging, vertical seismic profile, geological statistical data, or gravity data can be integrated with seismic data in a inversion framework, leveraging the complementary information from different data sources to improve the resolution and accuracy of the inversion results \cite{li2016integrated,li2021deep}.
n recent years, there has been spectacular advances in machine learning, especially deep learning (DL) \cite{ronneberger2015u,edinburgh31marchenko,vaswani2017attention,szegedy2017inception,lee2020maskgan,he2022masked}. Driven by the rapid development of network architecture, optimization algorithms and massive available data, DL has emerged as an important technique for tackling a variety of challenging tasks in computer vision (CV) and natural language processing (NLP), such as image classification, segmentation and speech recognition. DL has shown remarkable ability of learning the data features and building mappings between data domains. It is a fast-growing topic in geophysics, with rapidly-emerging application in seismic data processing, inversion and interpretation \cite{rasht2022}.

Low frequencies (LF) in seismic data are valuable for FWI. On one hand, LF reduce the reliance on starting models and mitigate the cycle-skipping issues. On the other hand, the LF penetrate the subsurface deeper and carry crucial wavefield information for reliable inversion. However, in practical exploration, seismic data are often band-limited, lacking low frequencies. Leveraging the strength of deep learning networks, many researchers have proposed DL-based approaches to extrapolate these low frequencies from bandwidth-limited observed seismic data \cite{ovcharenko2019deep,fang2020data,jin2021efficient,luo2023low}.

\begin{figure*} [htpb!]
	\centering
		\includegraphics[width=1\columnwidth]{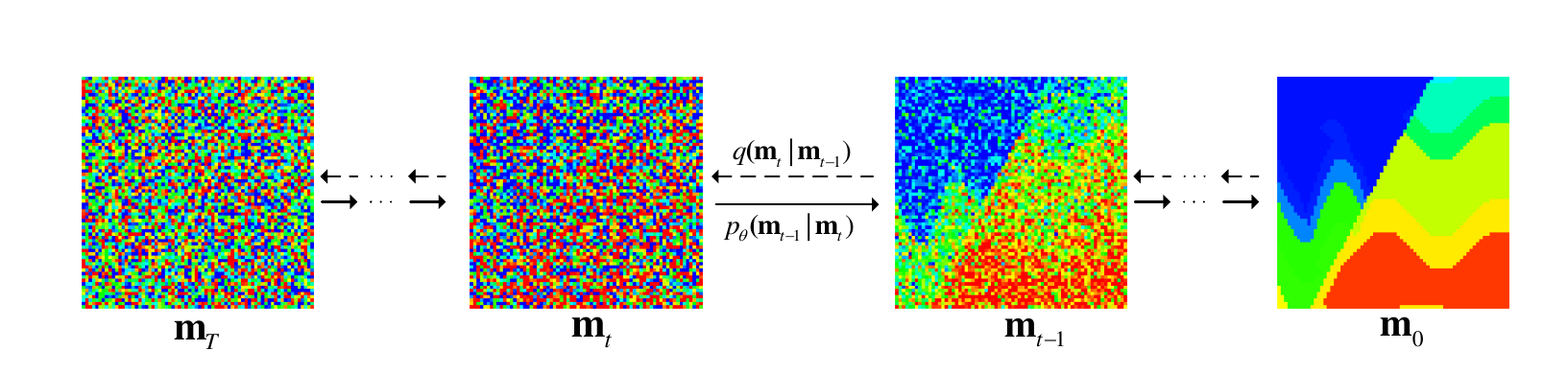}
 
\caption{An illustration of the forward (diffusion) and reverse (denoising) process in the unconditional GDM.}
 \label{fig:unconditional GDM}
\end{figure*}
As a valuable supplement to seismic data, prior information can help alleviate the ill-posed problem of FWI and improve the accuracy and reliability of the subsurface model. Various deep learning techniques have been developed to integrate prior information into the FWI process. \cite{zhang2019regularized} extract facies from well logs and use DNNs to build the mapping between the probability of facies and the inverted model. A prior model is then computed to regularize the elastic FWI. \cite{li2021deep} use the velocity distribution information, i.e., mean and variance, from wells as the features to obtain a high-resolution prior model. \cite{li2023self} introduce vision transformer (ViT) to build the network architecture and use the self-supervised pre-training and fine-tuning scheme to obtain the optimal mapping between seismic volume and well information. \cite{wang2023prior} incorporate the prior knowledge about the subsurface learned from a generative diffusion model (GDM) to regularize the FWI.

Given deep learning's remarkable capacity to handle complex non-linear operators, data-driven techniques have emerged for directly estimating the subsurface model parameters from observed seismic data. \cite{wu2019inversionnet} propose InversionNet, which is based on an encoder-decoder network with a conditional random field. VelocityGAN \cite{zhang2020data} uses a generative adversarial network (GAN) for velocity inversion and enhances generalization through the transfer learning strategy. \cite{zeng2021inversionnet3d} propose InversionNet3D to further extend InversionNet to a 3-D velocity inversion approach. One challenge for these data-driven techniques is that the DL networks trained on limited datasets often perform poorly on a new dataset. To mitigate this problem, \cite{wang2020velocity} and \cite{feng2021multiscale} augment the training datasets with velocity models style-transformed from natural images. 

 The GDMs, a state-of-the-art deep generative model, has the remarkable capability to learn important features in target data and generate representative samples, demonstrating superior performance in image and video generation \cite{ho2020denoising,dhariwal2021diffusion,ho2022video,rombach2022high}. Recently, the GDMs have been used to handle some geophysical tasks. \cite{wang2024reconstructing}, \cite{deng2024seismic}, and \cite{liu2024generative} use GDMs to reconstruct missing seismic traces. \cite{zhu2023diffusion}, \cite{zhang2024conditional}, and \cite{li2024conditional} use conditional GDMs to perform seismic signal separation and denoising. \cite{wang2023prior} use an unconditional GDMs to regularize the conventional FWI. Motivated by these work, we introduce the GDMs to the seismic velocity inversion and develop the DiffusionVel method that enables a reasonable integration of multiple information including seismic data, background velocity, prior geological knowledge, and well logs. In detail, we train two separate conditional GDMs with the seismic data and the well logs as the condition to predict the velocity model. These GDMs, named the seismic-data and well-log GDM, ensure the generated model can fit the constraints of seismic data and well logs, respectively. Additionally, we train an unconditional GDM, the geology-oriented GDM, on a distribution of velocity models to learn its geological features. The trained geology-oriented GDM can then adapt the generated velocity model to the learned prior geological knowledge. Besides, the background velocity is incorporated into the generated model by using a low-pass filter. We achieve the multi-information integration by incorporating these GDMs into the sampling process by a weighted summation. We can efficiently control the constraints of the multiple information by adjusting the weighting factors.
 
This paper is structured as follows. First, we provide a brief overview of the conventional FWI and the GDMs. Second, we introduce the methodology of the multi-information integration in DiffusionVel and outline the network architecture. Third, we evaluate the performance of the proposed DiffusionVel integrating multiple information including seismic data, background velocity, geological knowledge, and well logs by using the OpenFWI datasets and the Hess VTI model example. The DiffusionVel method is also compared with the conventional FWI, InversionNet, and VelocityGAN methods. Finally, we make a comprehensive discussion about the limitations and potential improvement for our DiffusionVel method.

\section{Theory}
\subsection{Conventional FWI}

Conventional FWI inverts for the subsurface model by minimizing the misfits between simulated and observed seismic data. The wave equation governing the seismic wave propagation is written as
\begin{equation} \label{eq:1}
\left[\frac{1}{K(\mathbf{x})}\frac{\partial^2}{\partial t^2}-\nabla\cdot\left(\frac{1}{\rho(\mathbf{x})}\nabla\right)\right]\mathbf{}u(\mathbf{x},t,\mathbf{x}_{s})=s(\mathbf{x},t,\mathbf{x}_{s}),
\end{equation}
where $\rho$ is the density, $K$ is the bulk modulus, $\mathbf{x}_s$ is the source location, $u$ is the pressure wavefield, and $s$ is the source signal.
Typically, the wave equation can be represented as
\begin{equation}\label{eq:2}
u=F(\mathbf{m}),
\end{equation}
where $\mathbf{m}$ is the subsurface model and $F$ is the forward modelling operator.
We can estimate the subsurface model by solving the following optimization problem:
\begin{equation}\label{eq:3}
\min_\mathbf{m}\left\{\left\|\mathbf{d}_{obs}-F(\mathbf{m})\right\|_2^2+\lambda
{R}(\mathbf{m})\right\},\end{equation}
where $\mathbf{d}_{obs}$ is the observed seismic data, $\lambda$ is the regularization weighting factor, and ${R}(\mathbf{m}) $ is the regularization term. 

Conventional FWI is a computationally intensive task that involves iterative forward modelling, gradient computation and optimization. Besides, it is challenging to predict the subsurface model accurately because of the inherent ill-posed problem. Here, we propose to use a data-driven method with generative diffusion model to address these limitations.

\subsection{Generative Diffusion Model }
We first review the theory of Generative Diffusion Model (GDM). You are referred to \cite{ho2020denoising} and \cite{zhang2024conditional} for more details. A graphic illustration of the unconditional GDM is shown in Figure~\ref{fig:unconditional GDM}. The GDMs first defines a forward (or diffusion) process that gradually introduces noise to a clean image $\mathbf{m}_0$ over T timesteps:
\begin{equation}\label{eq:4} 
\begin{aligned}
 q({\mathbf{m}}_{1:T}|{\mathbf{m}}_0)&=\prod\limits_{t=1}^T q({\mathbf{m}}_t|{\mathbf{m}}_{t-1}),
\\
q(\mathbf{m}_t|\mathbf{m}_{t-1})&=\mathcal{N}(\mathbf{m}_t;\sqrt{1-\beta_t}\mathbf{m}_{t-1},\beta_t\mathbf{I}),
\end{aligned}
\end{equation} 
where $\mathbf{m}_t$ is the noisy image at timestep $t$, and $\beta_t$ is the predefined noise-schedule that increases with $t$. At timestep $T$, $\mathbf{m}_t$ is corrupted into a pure noise image.
The property of the forward process allows
\begin{equation} \label{eq:5}
\mathbf{m}_t=\sqrt{\bar{\alpha}_t}\mathbf{m}_0+\sqrt{1-\bar{\alpha}_t}\boldsymbol{\epsilon},\quad\boldsymbol{\epsilon}\sim\mathcal{N}(\mathbf{0},\mathbf{I}),
\end{equation}
where $\alpha_t=1-\beta_t$. The GDMs then defines a reverse (denoising) process which should gradually remove the added noise from the pure noise image $\mathbf{m}_T\sim\mathcal{N}(\mathbf{0},\mathbf{I})$:
\begin{equation} \label{eq:6}
p_\theta(\mathbf{m}_{0:T})=p(\mathbf{m}_T)\prod\limits_{t=1}^Tp_\theta(\mathbf{m}_{t-1}|\mathbf{m}_t),
\end{equation}
However, the denoising operator $p_\theta(\mathbf{m}_{t-1}|\mathbf{m}_t)$ is unknown. In other words, the added noise in every forward step is unknown when we take the reverse step. Therefore, the GDMs learn to predict the noise by optimizing the loss function with a DNN $\theta$: 
\begin{equation} \label{eq:7}
\mathbb{E}_{t,\mathbf{m}_t,\boldsymbol{\epsilon}}\Big[\left\|\boldsymbol{\epsilon}-\boldsymbol{\epsilon}_{\theta}(\mathbf{m}_{t},t)\right\|^2\Big].
\end{equation} 
Once the training process of GDMs is finished, GDMs can generate clean images in the sampling process. One challenge for the traditional sampling process is the computational cost brought by numerous timesteps. Here, we adopt the diffusion denoising implicit model (DDIM) \cite{song2020denoising} to reduce the sampling timesteps.  Every sampling step consists of two steps :
\begin{equation}\label{eq:8}
\begin{aligned}
\mathbf{m}_{0_{\theta}}(\tau)=\frac{\mathbf{m}_{\tau}-\sqrt{1-\alpha_{\tau}}\boldsymbol{\epsilon}_{\theta}(\mathbf{m}_{\tau},\tau)}{\sqrt{\alpha_{\tau}}},
\end{aligned}
\end{equation}

\begin{equation}\label{eq:9}
\begin{aligned}
\mathbf{m}_{\tau-1}=\sqrt{\alpha_{\tau-1}}\mathbf{m}_{0_{\theta}}(\tau)+&\sqrt{1-\alpha_{\tau-1}-\sigma_{\tau}^{2}}\boldsymbol{\epsilon}_{\theta}(\mathbf{m}_{\tau},\tau)+\sigma_{\tau}\boldsymbol{\epsilon},
\\
&\boldsymbol{\epsilon}\sim\mathcal{N}(\mathbf{0},\mathbf{I}),
\end{aligned}
\end{equation}
where $\tau$ is the timestep sampled from a  sub-sequence of original time sequence 0 ... $T$,  and $\sigma_{\tau}=\sqrt{(1-\alpha_{\tau-1})/(1-\alpha_{\tau})}\sqrt{(1-\alpha_{\tau}/\alpha_{\tau-1})}$, $\mathbf{m}_{0_{\theta}}$ is the estimated clean image.  For clarity we will refer to Equation \ref{eq:8} as the unconditional prediction step $G_{\theta}(\mathbf{m}_{\tau},\tau)$ since it predicts the clean image for the current time step.  Equation \ref{eq:9} will be referred to as the diffusion step $D(\mathbf{m}_{0_{\theta}}(\tau),\tau)$ , as it introduces noise to the estimated clean image to generate the noisy image input for the next sampling step.

The unsupervised nature of unconditional GDMs enables effective learning of the target distribution, but their generation is random. Thus, we further introduce the conditional GDM. The condition ${y}$ in the conditional GDM serve as a guidance to the sampling process of GDM. The modified loss function is 
\begin{equation} \label{eq:10} 
\mathbb{E}_{t,\mathbf{m}_t,\mathbf{y},\boldsymbol{\epsilon}}\Big[\left\|\boldsymbol{\epsilon}-\boldsymbol{\epsilon}_{\theta}(\mathbf{m}_{t},\mathbf{y},t)\right\|^2\Big].
\end{equation}

The sampling step in Equation \ref{eq:8} and Equation \ref{eq:9} becomes

\begin{equation}\label{eq:11}
\begin{aligned}
\mathbf{m}_{0_{\theta}}(\tau)=\frac{\mathbf{m}_{\tau}-\sqrt{1-\alpha_{\tau}}\boldsymbol{\epsilon}_{\theta}(\mathbf{m}_{\tau},\mathbf{y},\tau)}{\sqrt{\alpha_{\tau}}},
\end{aligned}
\end{equation}

\begin{equation}\label{eq:12}
\begin{aligned}
\mathbf{m}_{\tau-1}=\sqrt{\alpha_{\tau-1}}\mathbf{m}_{0_{\theta}}(\tau)+&\sqrt{1-\alpha_{\tau-1}-\sigma_{\tau}^{2}}\boldsymbol{\epsilon}_{\theta}(\mathbf{m}_{\tau},\mathbf{y},\tau)+\sigma_{\tau}\boldsymbol{\epsilon},
\\
&\boldsymbol{\epsilon}\sim\mathcal{N}(\mathbf{0},\mathbf{I}).
\end{aligned}
\end{equation}
Thus, Equation \ref{eq:11} is the conditional prediction step $G_{\theta}(\mathbf{m}_{\tau},\mathbf{y},\tau)$  and  Equation \ref{eq:12} is  the diffusion step $D(\mathbf{m}_{0_{\theta}}(\tau),\tau)$ .

\begin{figure*} [htpb!]
	\centering
	
	\includegraphics[width=1\columnwidth]{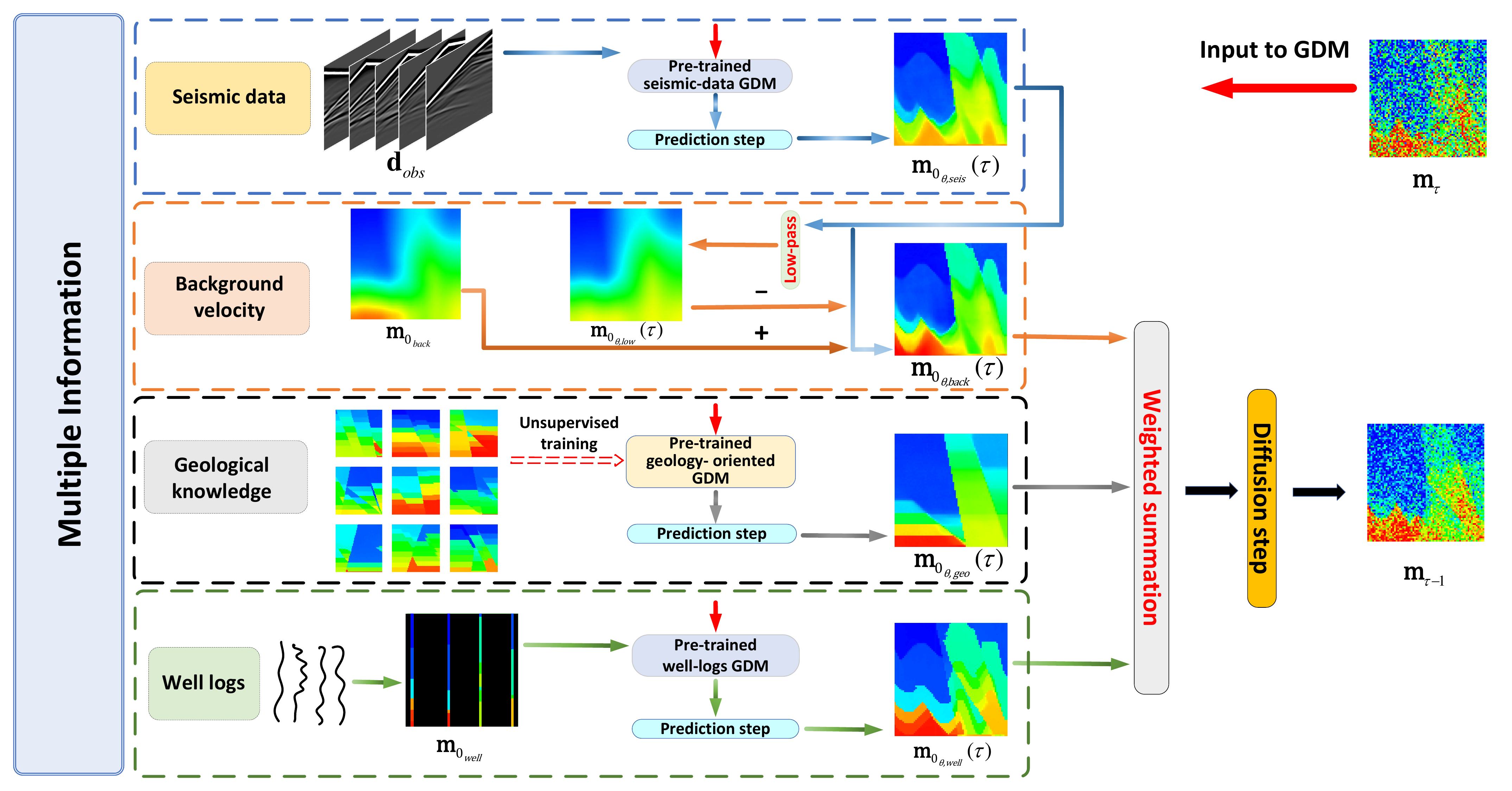}
\caption{A graphic illustration of the multi-information integration in DiffusionVel. We use two separate pre-trained conditional GDMs for seismic data and well logs, respectively. The condition for the conditional GDMs is observed seismic data or well-log image. We use a low-pass filter to extract the LF information from the velocity model estimated by the seismic GDM and replace it with the given background velocity model. We pre-train an unconditional GDM on a velocity model distribution that is consistent with our prior geological knowledge of the subsurface in an unsupervised manner. Each GDM (conditional or unconditional ) uses the noisy velocity model $\mathbf{m}_{\tau}$ at the current sampling step as input and outputs the estimated velocities with corresponding constraints using the prediction step in Equation \ref{eq:8} or Equation \ref{eq:11}. Finally, these estimates are combined together through a weighted summation and then use one diffusion step in Equation \ref{eq:12} to produce the noisy velocity model $\mathbf{m}_{\tau-1}$ of next sampling step.}

\label{fig: muti-information integration}
\end{figure*}
\subsection{DiffusionVel}
The proposed DiffusionVel integrates multiple information including seismic data, background velocity, geological knowledge and well logs to predict the subsurface model by using the GDMs. A graphic illustration of the DiffusionVel incorporating the multiple information is shown in Figure~\ref{fig: muti-information integration}. The four information is incorporated into sampling process by using (1) a pre-trained conditional GDM for the seismic data, named as the seismic-data GDM, (2) a low-pass filter for the background velocity, (3) a pre-trained unconditional GDM for the prior geological knowledge, named as the geology-oriented GDM, and (4) a pre-trained conditional GDM for the well logs, named as the well-log GDM. Among these information, the seismic data is used as the main data source during the velocity inversion while the others serve as prior information to regularize the sampling process. 
\\
\subsubsection{Seismic data}

We can establish the mapping from observed seismic data to velocity model by using them as the condition $\mathbf{y}$ and the target data distribution $\mathbf{m}_0$ of the conditional GDM, respectively. The seismic-data GDM, represented as $_{\theta, seis}$,  is viewed as a data-driven method. The data-driven methods use a neural network to approximate the inverse operator $F^{-1}$ directly, and thus generate the velocity model corresponding to the seismic data. According to the conditional prediction step in Equation \ref{eq:11}, the velocity model estimated by $\theta_{seis}$ at timestep $\tau$ is
\begin{equation}\label{eq:13}
\mathbf{m}_{0_{\theta, seis}}(\tau)=G_{\theta, seis}(\mathbf{m}_{\tau},\mathbf{d}_{obs},\tau)=F^{-1}(\mathbf{d}_{obs}).
\end{equation}

\subsubsection{Background velocity}

It is challenging for the neural network to learn such a highly-nonlinear mapping relations from seismic data to velocity models, especially when the velocity models exhibit complex structures. \cite{choi2021ilvr} use LF contents of reference image to constrain the sampling process. Considering that the background velocity model is usually available, we can incorporate the background model into every sampling step to constrain the velocity estimation. 

 We assume that background velocities represent the LF components filtered from the true models $\mathbf{m}_{0}$:
\begin{equation}\label{eq:14}
  \mathbf{m}_{0_{back}}=f_{low}(\mathbf{m}_{0}),
\end{equation}
where $f_{low}$ is a low-pass filter. This low-pass filter is also applied to the seismic-estimated velocity model  in every sampling step to extract its LF components, represented as
\begin{equation}\label{eq:15}
 \mathbf{m}_{0_{\theta, low}}(\tau)=f_{low}(\mathbf{m}_{0_{\theta, seis}}(\tau)).
\end{equation} 
We compute the model residuals between $\mathbf{m}_{0_{\theta, low}}(\tau)$ and $\mathbf{m}_{0_{back}}$ and then sum them with the seismic-estimated velocity model $\mathbf{m}_{0_{\theta,seis}}(\tau)$. The corresponding formula for obtaining the background-integrated velocity model is 
\begin{equation}\label{eq:16}
\mathbf{m}_{0_{\theta, back}}(\tau)=\mathbf{m}_{0_{\theta, seis}}(\tau)+\mathbf{m}_{0_{back}}- \mathbf{m}_{0_{\theta, low}}(\tau).
\end{equation}

\subsubsection{Geological knowledge}

The seismic-data GDM learns and samples from the velocity model distribution under the supervision of seismic data. Consequently, the seismic-data GDM trained on a set of data sharing similar geological features tends to generate the model that represents the similar features. For example, when applying the seismic-data GDM trained on curve-layered velocity models to the seismic data produced by flat-layered velocity model, the layers of the generated models appear curved rather than flat. To generalize well on all datasets, one feasible way is to include these datasets during training. However, it is computationally expensive to prepare various velocity model and simulate the seismic data for each velocity model. To mitigate this, we can use a velocity model distribution with prior geological knowledge to train the unconditional GDM. Then the GDM with learned geological knowledge can be combined with the seismic-data GDM. With the integration of background velocity shown in Equation \ref{eq:16}, the generated velocity model in every sampling step can be written as 
\begin{equation}\label{eq:17}
\mathbf{m}_{0_{\theta, integration}}(\tau)=\lambda_{1}\mathbf{m}_{0_{\theta, back}}(\tau)+(1-\lambda_{1})\mathbf{m}_{0_{\theta ,geo}}({\tau}),
\end{equation}
where the subscript $_{\theta, geo}$ represents the GDM with learned prior geological knowledge, $\mathbf{m}_{0_{\theta, geo}}(\tau)=G_{\theta, geo}(\mathbf{m}_{\tau},\tau)$ is the velocity model estimated by the unconditinoal prediction step in Equaiton \ref{eq:8} and $\lambda_1$ is a weighting factor between 0 and 1. When $\lambda_1=0$ (or 1), only seismic data (or geological knowledge) contributes to the sampling process. We can easily adjust the weighting factor to balance their contribution.  Without loss of generality, we denote the multi-information (besides seismic data and background velocity) integrated velocity model  at every sampling step as $ \mathbf{m}_{0_{\theta, integration}}(\tau)$ in the rest of the paper.

\subsubsection{Well information}

Well logs can provide high-resolution information of the subsurface properties, but only sample limited locations in the lateral direction. We train a conditional GDM by using the well-velocity image and the velocity model as the condition and target of the generative model. We produce the well-velocity image by directly inserting the well velocity at well locations of the image while keep the rest of the image to zero. After the training process, the well-log GDM can generate the velocity model constrained by the well information given sampled well velocity. Then,  with the integration of background velocity shown in Equation \ref{eq:16}, we combine the generated velocity model constrained by the well information with the generated velocity model driven by seismic data with a weighted summation in each sampling step: 
\begin{equation}\label{eq:18}
\mathbf{m}_{0_{\theta, integration}}=\lambda_{2}\mathbf{m}_{0_{\theta, back}}(\tau)+(1-\lambda_{2})\mathbf{m}_{0_{\theta ,well}}({\tau}),
\end{equation}
where the subscript $_{\theta, well}$ represents the well-log GDM, $\mathbf{m}_{0_{\theta ,well}}({\tau})=G_{\theta, well}(\mathbf{m}_{\tau},\mathbf{m}_{0_{well}},\tau)$ is the velocity model estimated by the conditional prediction step in Equaiton \ref{eq:11} with $\mathbf{m}_{0_{well}}$ being the well-log image, and $\lambda_2$ is a weighting factor. The influence of $\lambda_2$ on the final inversion result will be investigated in the experiment section.

\subsection{Network design}

\begin{figure} [htpb!]
	\centering
	\includegraphics[width=0.7\columnwidth]{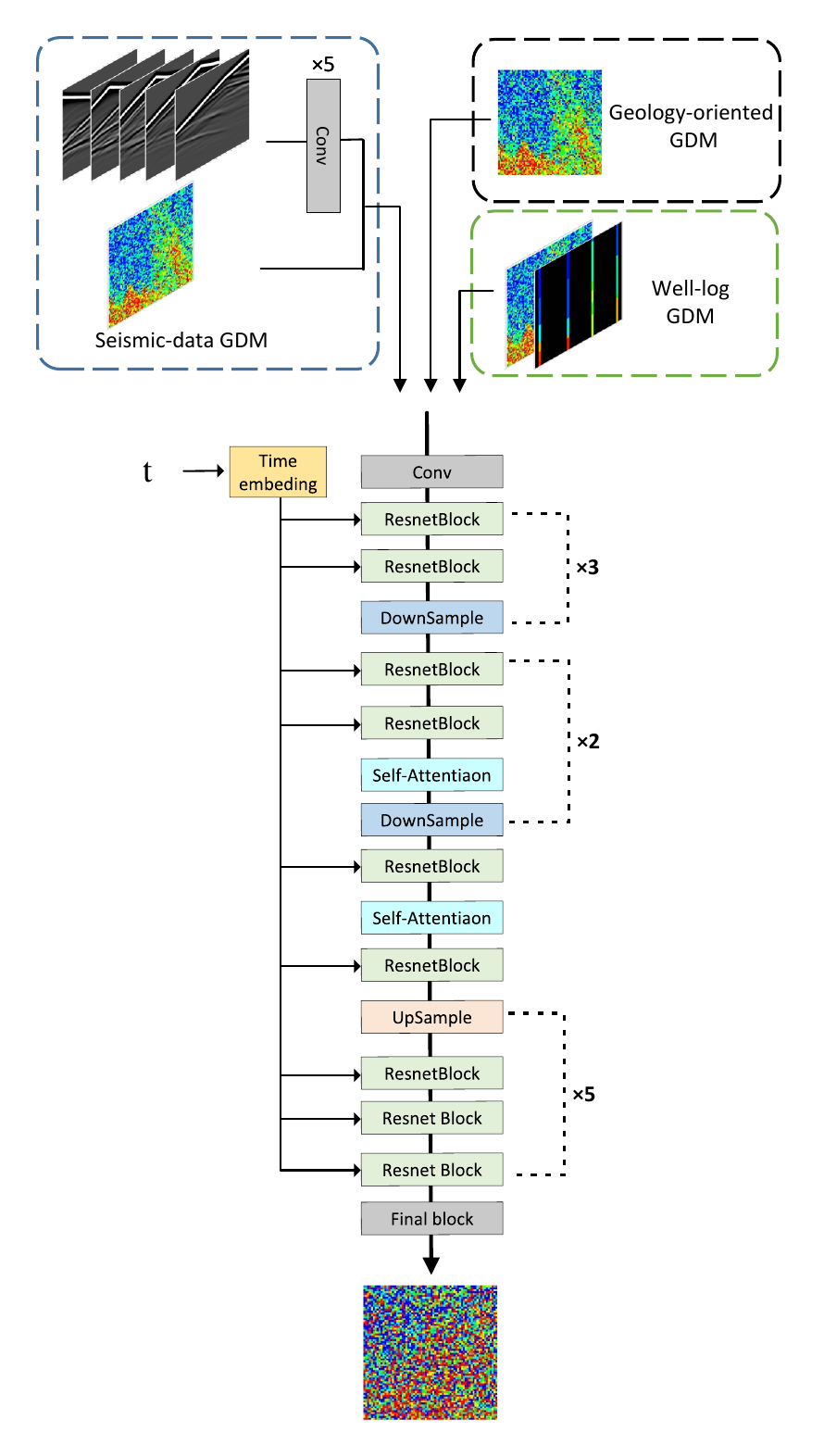}  
\caption{A graphic illustrate of the U-net structure shared by the seismic-data GDM, the geology-oriented GDM and the well-log GDM. The skip connection for U-net is not shown. }
 \label{fig:U-net arch}
\end{figure}

These three GDMs corresponding to seismic data, well information and prior geological knowledge use the same U-net structure shown in Figure~\ref{fig:U-net arch} , a legacy backbone archtecture from our preivous research  \cite{zhang2024conditional}.  The U-net of the unconditional GDM for learning prior geological knowledge use the noisy velocity model and the timestep as the input, and use noise image as the output. For the GDMs with seismic data or well information as condition, the input of the U-net consists of the noisy velocity model, timestep, and seismic data or well-log image and the target output is noise image. Before concatenating the seismic data with the noisy velocity model, First, we pad the seismic data with 105 zeros along the travel-time direction. Then, we apply four convolutional layers to reduce the dimension of the seismic data to match that of the noisy velocity model. The kernel size, stride, and padding for all four convolutional layers are 3×3, 2×1, and 1×1, respectively. We introduce the residual modules and the multi-head self-attention mechanism into the U-net to enhance the performance of the network. The timestep is used as the input of the U-net through a time-embedding block that includes the positional encoding \cite{vaswani2017attention}. One can find more  details of the used U-net architecture in \cite{zhang2024conditional}. The MSE loss  is used as the loss function for training the network. 

\begin{figure*} [htpb!]
	\centering
 {
 \includegraphics[width=1\columnwidth]{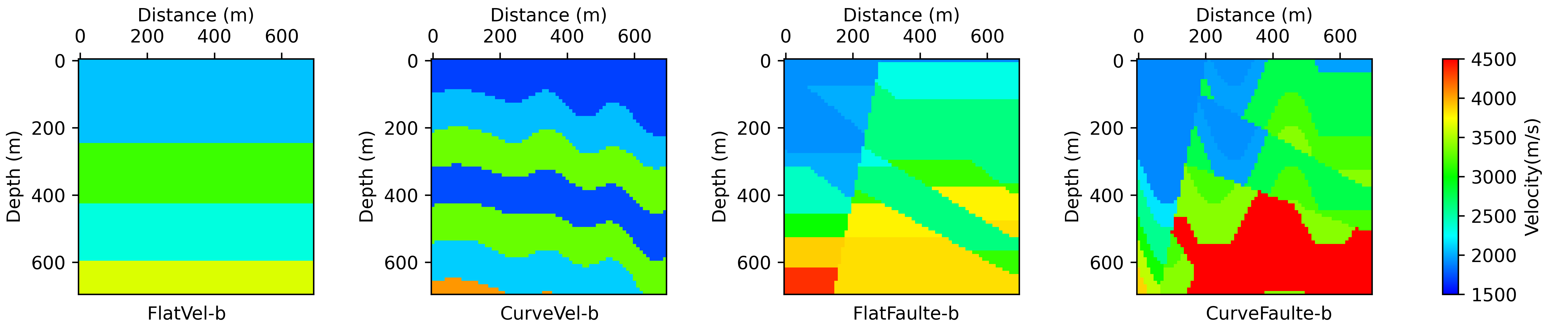}}

\caption{Examples of velocity models in the OpenFWI datasets. }
 \label{fig:datasets_example_velocities}
\end{figure*}
\begin{figure*} [htpb!]
	\centering
	{
		\includegraphics[width=1\columnwidth]{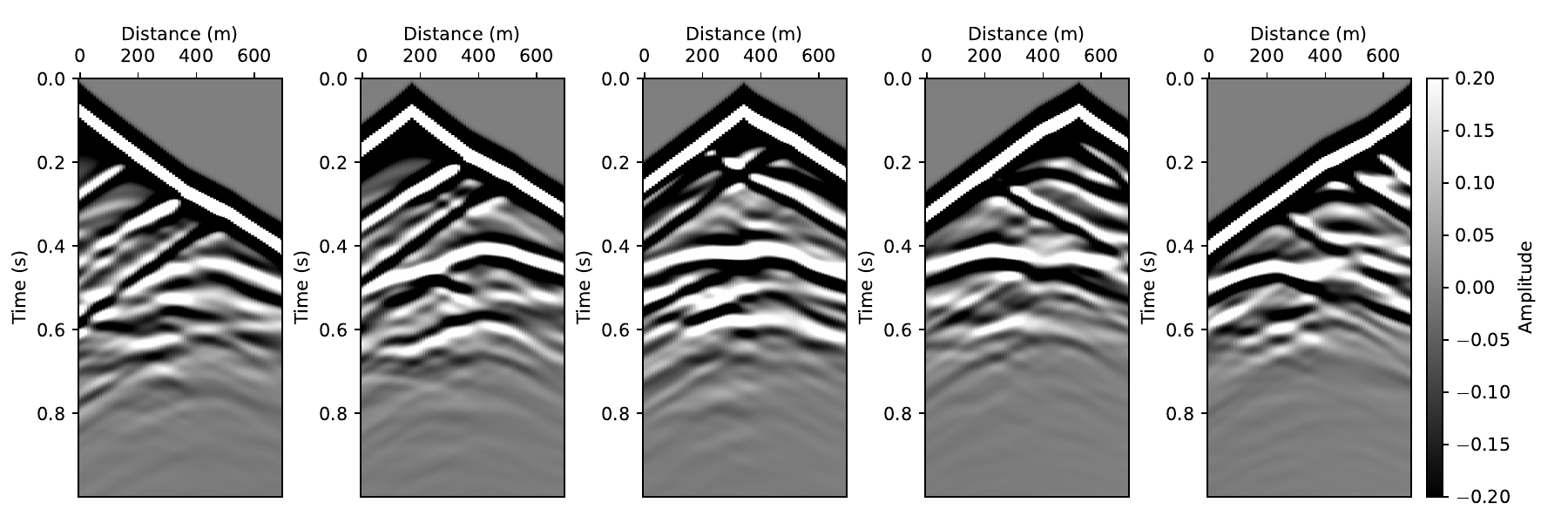}}
  
\caption{Five shot gathers simulated with the CurveFault-B velocity model in Figure~\ref{fig:datasets_example_velocities}.}
 \label{fig:datasets_example_data}
\end{figure*}
\section{Experiments }
\subsection{Experiments Settings}\label{subsec: experiments_setting}

The data-driven DL methods require a large amount of high-quality data to train the network to learn a well-generalized mappings between two data spaces. Here, we use the OpenFWI datasets \cite{OpenFWIdeng2022} to train our proposed DL method. We also compare our proposed method with the previously developed DL-based seismic velocity inversion methods, InversionNet and VelocityGAN, by using the OpenFWI datasets. These datasets include velocity models exhibiting diverse subsurface structures, along with simulated seismic data. Among these datasets, we use the "FlatVel-B", "CurveVel-B", "FlatFault-B", and "CurveFault-B" datasets for training. Figure~\ref{fig:datasets_example_velocities} shows one velocity model example from each of these datasets. Figure~\ref{fig:datasets_example_data} shows the simulated seismic data with five shot gathers corresponding to the CurveFault-B velocity model in Figure~\ref{fig:datasets_example_velocities}. The detailed description of the selected datasets is shown in Table~\ref{tab:datasets}. A total of five shots with an interval of 175 m are used for each velocity model to simulate its seismic data. There are 70 receivers to record the seismic wave. A 15-Hz Ricker wavelet is used as the source wavelet. For a fair comparison, we use the same training dataset and the same MSE loss function to train the InversionNet, the VelocityGAN and the proposed GDM driven by seismic data. We train the GDM using an Adam optimizer with an initial learning rate of 1e-4. The batch size is eight, and the maximum training epoch is set to 200. We apply the cosine noise schedule strategy proposed by \cite{nichol2021improved} to control \(\beta_t\) in Equation \ref{eq:4}. The total training timesteps is 1000. We use five sampling steps to generate the velocity model in the seismic-data example. In the other examples, we use 20 sampling steps to obtain better generation. 
\begin{table}[b]
  \centering
  \caption{Description of the OpenFWI datasets}
\label{tab:datasets}
   \resizebox{\textwidth}{!}{
\begin{tabular}{|l|l|l|l|l|}
\hline
Datasets & Description & Training/Testing & Seismic Data Size & Velocity Map Size \\
\hline
FlatVel-B & Flat layers & 23.5k/1k& $5\times1000\times1\times70$ & $70\times1\times70$ \\
\hline
CurveVel-B & Curved layers & 21.5k/0.5k& $5\times1000\times1\times70$ & $70\times1\times70$ \\
\hline
Flatfault-B & Flat layers with multiple faults & 24k/3k& $5\times1000\times1\times70$ & $70\times1\times70$ \\
\hline
Curvefault-B & Curved layers with multiple faults & 22.5k/1.5k& $5\times1000\times1\times70$ & $70\times1\times70$ \\
\hline
\end{tabular}
}
\end{table}

\begin{figure*} [htpb!]
	\centering
	{\includegraphics[width=1\columnwidth]{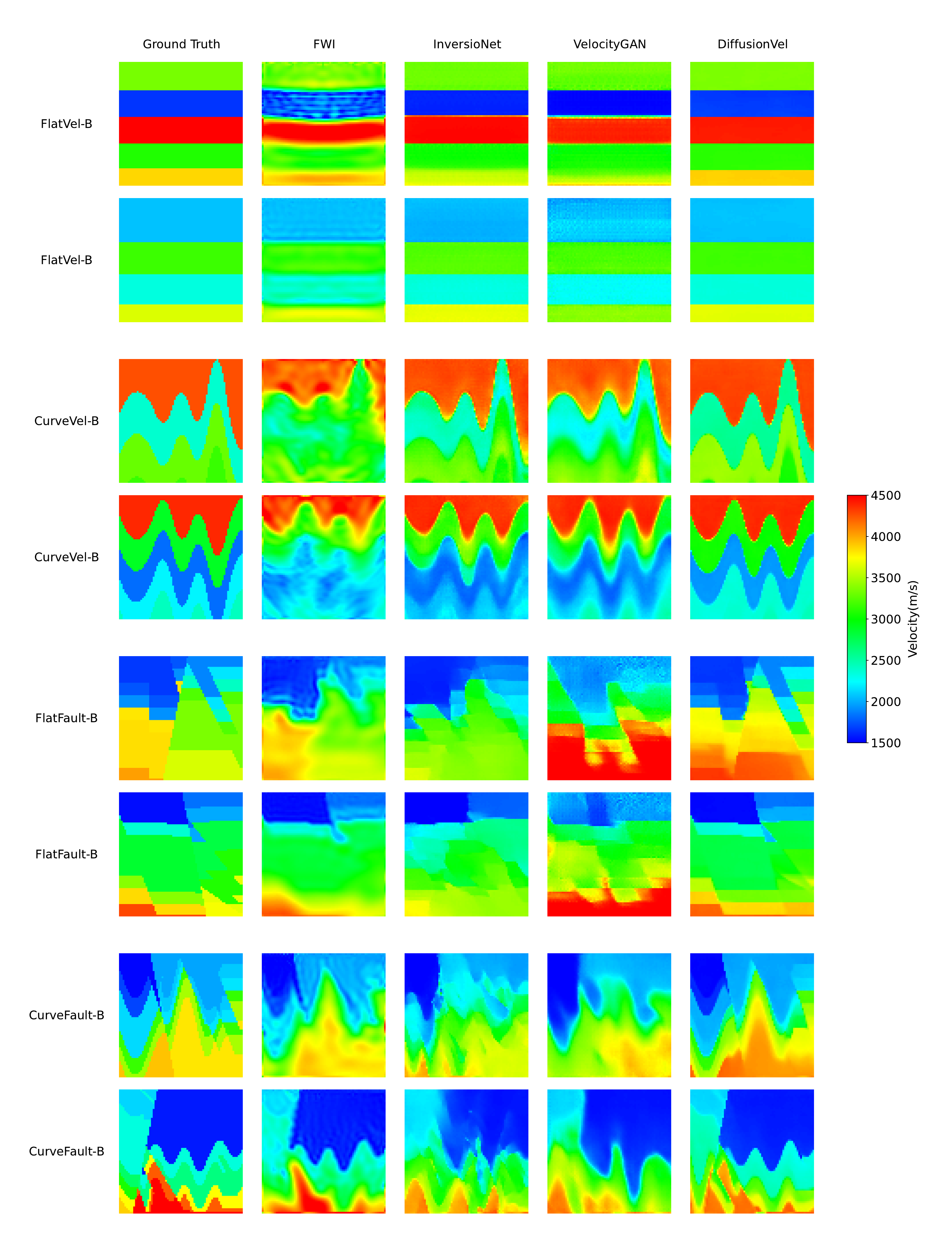}}
\caption{Inversion results for two model example respectively from the FlatVel-B, CurveVel-B, FlatFault-B and CurveFault-B datasets. The five columns from left to right show the ground-truth model, conventional FWI result, InversionNet result, velocityGAN result, and the seismic-data GDM result, respectively.}
 \label{fig:part1}

\end{figure*}
 \begin{figure*} [htpb!]
	\centering
	{
	\includegraphics[width=1\textwidth]{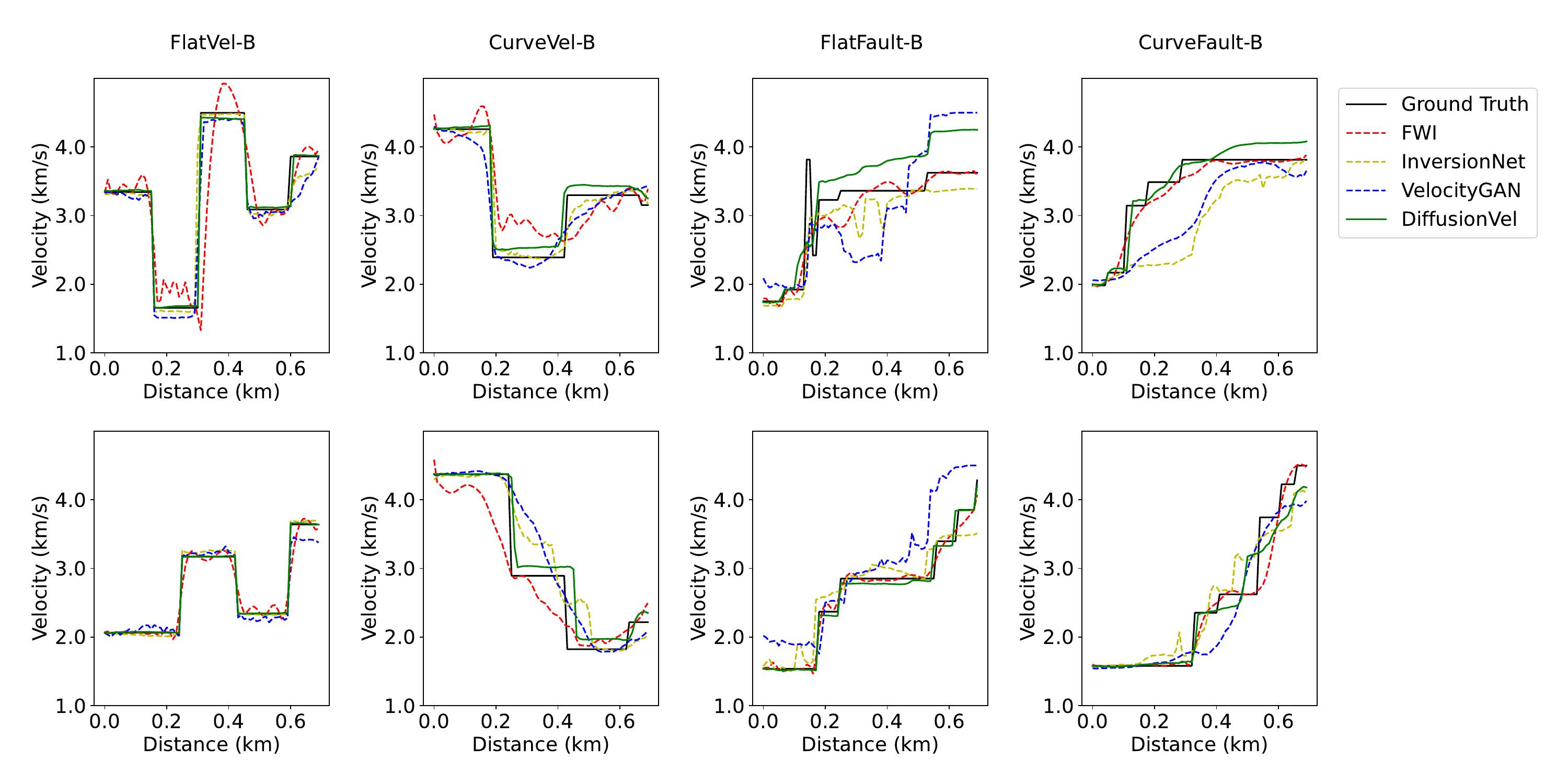}}
 
\caption{The vertical velocity profiles at the center of generated models in Figure~\ref{fig:part1}.}
\label{fig:part1_single_trace}
\end{figure*}
\begin{table}[b]
 \centering
 \caption{The quality metrics for different inversion methods}
  \label{tab:metrics1}
  \resizebox{\textwidth}{!}{
  \begin{tabular}{|c|ccc|ccc|ccc|ccc|}
    \hline
    Datasets & \multicolumn{3}{c|}{FWI} & \multicolumn{3}{c|}{InversionNet} & \multicolumn{3}{c|}{VelocityGAN} & \multicolumn{3}{c|}{DiffusionVel} \\
    \cline{2-13}
    & MAE & MSE & SSIM & MAE & MSE & SSIM & MAE & MSE & SSIM & MAE & MSE & SSIM \\
    \hline
    FlatVel-B & 0.0846 & 0.0262 &  0.7580 & 0.0437 & 0.0092 &  0.9343 & 0.0700 & 0.0147 & 0.8713 & \textbf{0.0232} & \textbf{0.0044} & \textbf{0.9738} \\
    \hline
    CurveVel-B & 0.1760 & 0.0662 & 0.6019 & 0.1695 & 0.0888 & 0.6604 & 0.1651 & 0.0759 & 0.6659 & \textbf{0.0983} & \textbf{0.0393} & \textbf{0.8331} \\
    \hline
    FlatFault-B & \textbf{0.0567} & \textbf{0.0094} & 0.8162 & 0.2528 & 0.1172 & 0.5890 & 0.2370 & 0.1009 & 0.5248 & 0.0714 & 0.0172 & \textbf{0.8448} \\
    \hline
    CurveFault-B & \textbf{0.0800} & \textbf{0.0170} & \textbf{0.7689} & 0.2424 & 0.1087 & 0.5918 & 0.2252 & 0.0977 & 0.5611 & 0.1280 & 0.0420 & 0.7094 \\
    \hline
  \end{tabular}
  }
\end{table}

\subsection{The GDM driven by seismic data}\label{sec:seismic-data}
We first apply the seismic-data GDM in our DiffusionVel to the FlatVel-B, CurveVel-B, FlatFault-B and CurveFault-B datasets, respectively. The first and last column of Figure~\ref{fig:part1} shows the true velocity models and the generated velocity models by using the different datasets to train the seismic-data GDM, respectively. We can see that the generated models look close to the true models in terms of the structure, and the sharp interfaces are well preserved with high resolution. However, the velocity values in some deep layers are not recovered accurately.

We then make a detailed comparison between our proposed method and the conventional FWI, InversionNet and VelocityGAN methods by using the same datasets. We smooth the true velocity model by using a Gaussian filter with a kernel size of 25 to obtain the initial model for the conventional FWI method. The final FWI result is shown in the second column in Figure~\ref{fig:part1}. We can see that the velocity models are somewhat recovered, but the interfaces are obviously smeared. The third and fourth columns show the velocity models generated by the InversionNet and VelocityGAN methods, respectively. In the FlatVel-b and CurveVel-b examples, the velocity models generated by the InversionNet and VelocityGAN is comparable to the true ones. However, the InversionNet and VelocityGAN did not perform well in the FlatFault-b and CurveFault-b examples. Intuitively, DiffusionVel generate the velocity model with highest resolution and accuracy. Figure~\ref{fig:part1_single_trace} shows the vertical velocity profiles at the center of generated models in Figure~\ref{fig:part1}.

We calculate the quality metrics for these inversion methods, using the mean absolute error (MAE), mean square error (MSE), and structural similarity index measure (SSIM) to evaluate the difference between the true velocity model and the generated velocity model. Lower values of MAE and MSE, along with higher values of SSIM, indicate a closer resemblance of the generated velocity model to the true velocity model. The quality metrics of these four methods is presented in Table~\ref{tab:metrics1}. By analyzing Table~\ref{tab:metrics1}, we find that (1) DiffusionVel generates the velocity model with highest quality compared with the other three methods on FlatVel-B and CurveVel-B datasets; (2) conventional FWI estimates velocity models of slightly better quality than DiffusionVel. This can be attributed to the background velocity that is only used in the conventional method.

\subsection{Integration of background velocity}

\begin{figure*} [htpb!]
	\centering
	{
		\includegraphics[width=1\columnwidth]{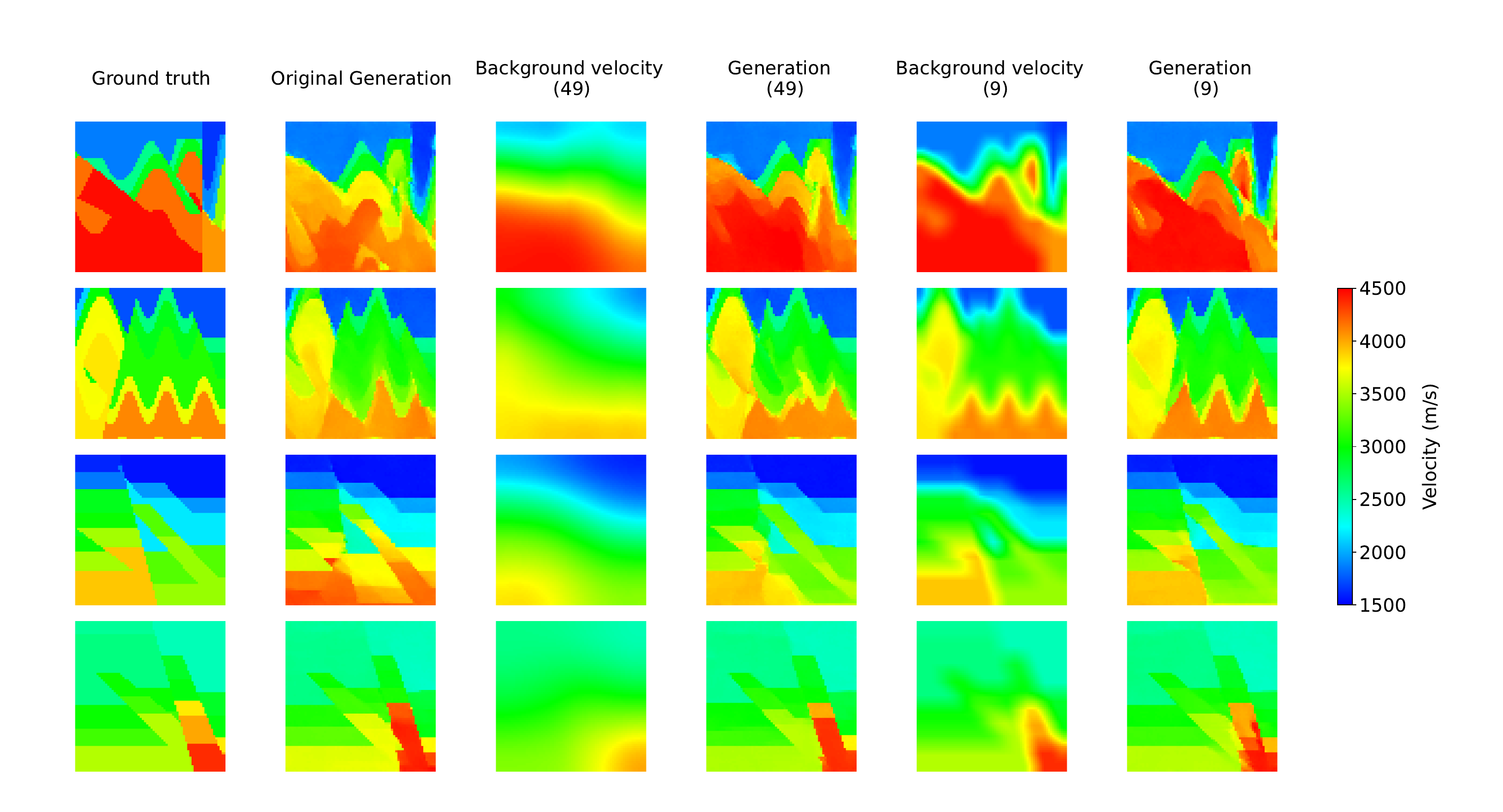}}
 
\caption{The generation results of the seismic-data GDM by using the integrated background model. From left to right: the ground-truth models, the original generation of seismic-data GDM, the background velocity smoothed with a Gaussian kernel of 49, the generation when the background models shown in the third column is integrated into the GDM, the background velocity smoothed with a Gaussian kernel of 9, the generation when the background models shown in the fifth column is integrated into the GDM.}
 \label{fig:datasets_lowf}
\end{figure*}

We then test how the background velocity integrated into the sampling process influence the generation result. We first prepare the background model by smoothing the true model with the Gaussian filter. The kernel size of the Gaussian filter is used to control the smoothness of the background model. Table~\ref{tab:background} shows the quantitative comparison between the integration of background velocity using three different kernel sizes. The table indicates that the integrated background model help to improve the generation accuracy, and the accuracy of the generated model increase with that of the incorporated background velocity.
\begin{table}[b]
\centering
\caption{Quantitative results of using different kernel size}
\label{tab:background}
\resizebox{\textwidth}{!}{
\begin{tabular}{|c|ccc|ccc|ccc|}
\hline
Datasets & \multicolumn{3}{c|}{49} & \multicolumn{3}{c|}{19} & \multicolumn{3}{c|}{9} \\
\cline{2-10}
& MAE & MSE & SSIM & MAE & MSE & SSIM & MAE & MSE & SSIM \\
\hline
FlatFault-B & 0.0481 & 0.0122 & 0.8461 & 0.0426 & 0.0106 & 0.8537 & 0.0335 & 0.0076 & 0.8820 \\
\hline
CurveFault-B & 0.0987 & 0.0300 & 0.7060 & 0.0925 & 0.0288 & 0.7027 & 0.0643& 0.0167& 0.7876\\
\hline
\end{tabular}
}
\end{table}
Figure~\ref{fig:datasets_lowf} displays four examples from the FlatFault-B and CurveFault-B datasets of integrating the background velocity. The first and second columns show the true velocity model and the generated velocity model without background-velocity integration. We can see that the velocity values in the deep layers are not reconstructed accurately. The third and fourth columns show the background model smoothed with a Gaussian kernel of 49 and the generated velocity model by integrating the background models. We can see that the integrated background model helps to improve the generation accuracy, especially in the deep area. The fifth and sixth columns show the background model smoothed with the Gaussian kernel of 9 and the generated velocity model by integrating the background models. We can see that a finer background velocity, offering more information on the velocities, can further enhance the inversion result. 

\subsection{Integration of geological knowledge}
\begin{figure*}[htpb!]
    \centering
 
    \subfloat[]{\includegraphics[width=0.45\columnwidth]{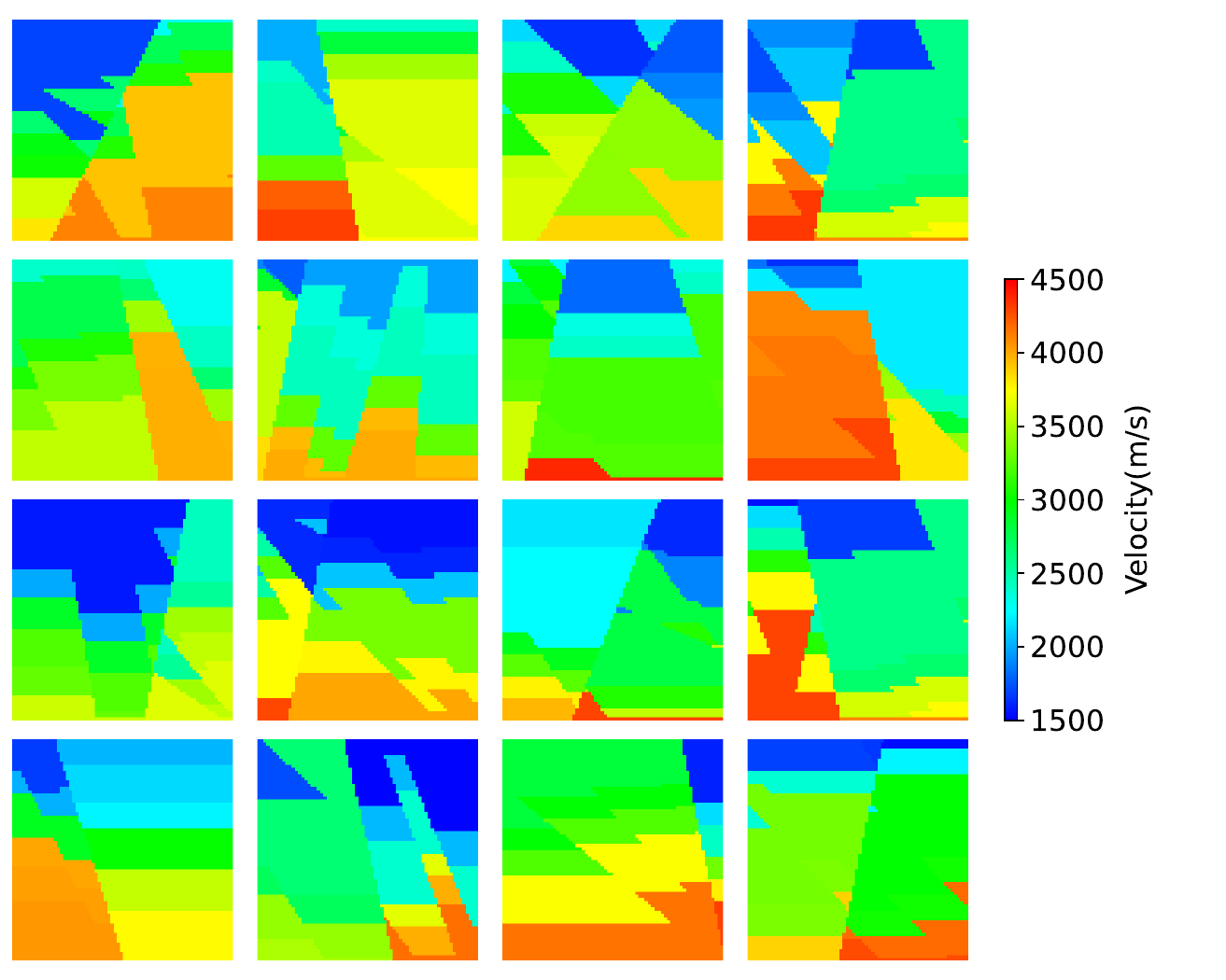}}\quad
    \subfloat[]{\includegraphics[width=0.45\columnwidth]{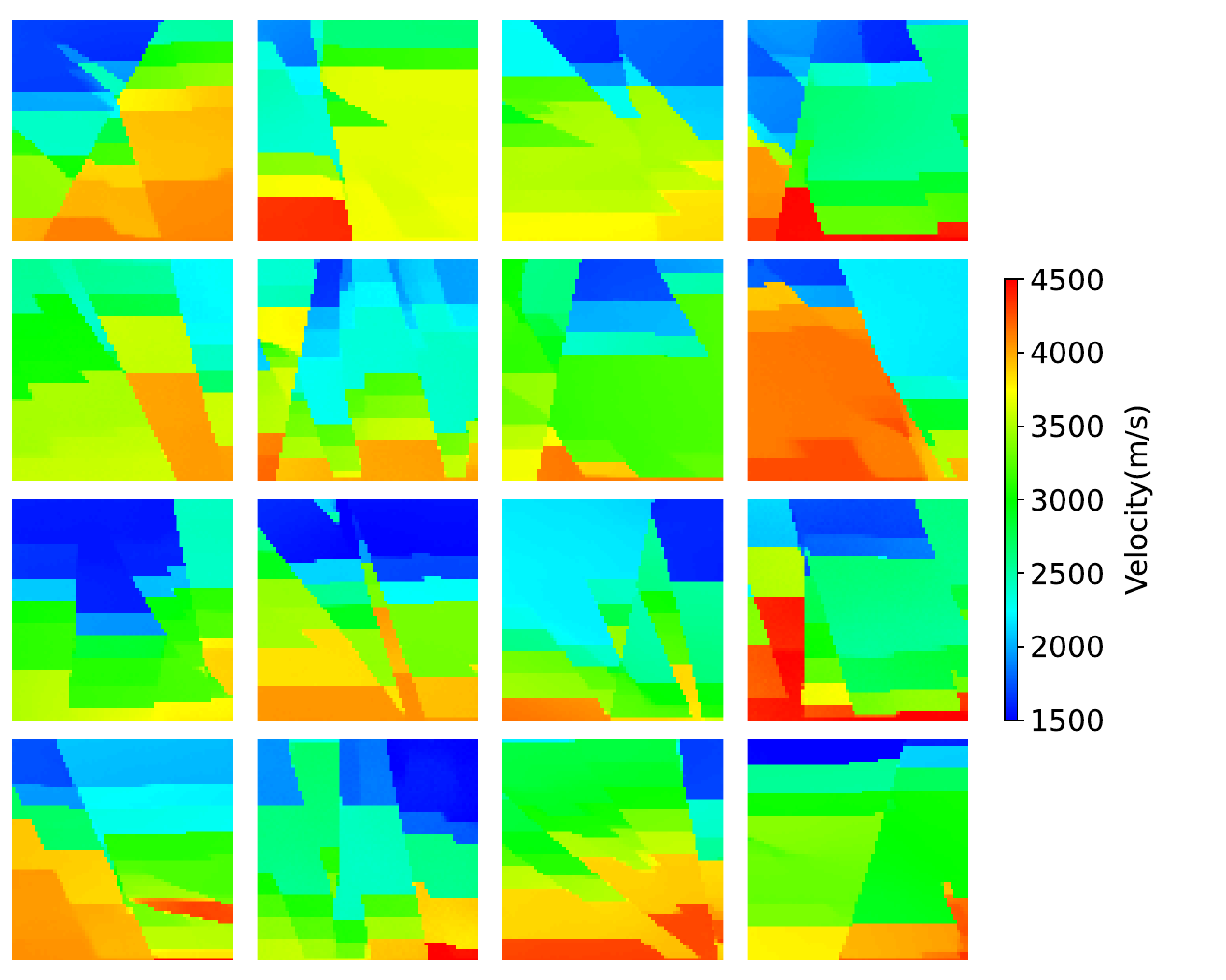}\label{fig:geo_0}}\\
    \subfloat[]{\includegraphics[width=0.45\columnwidth]{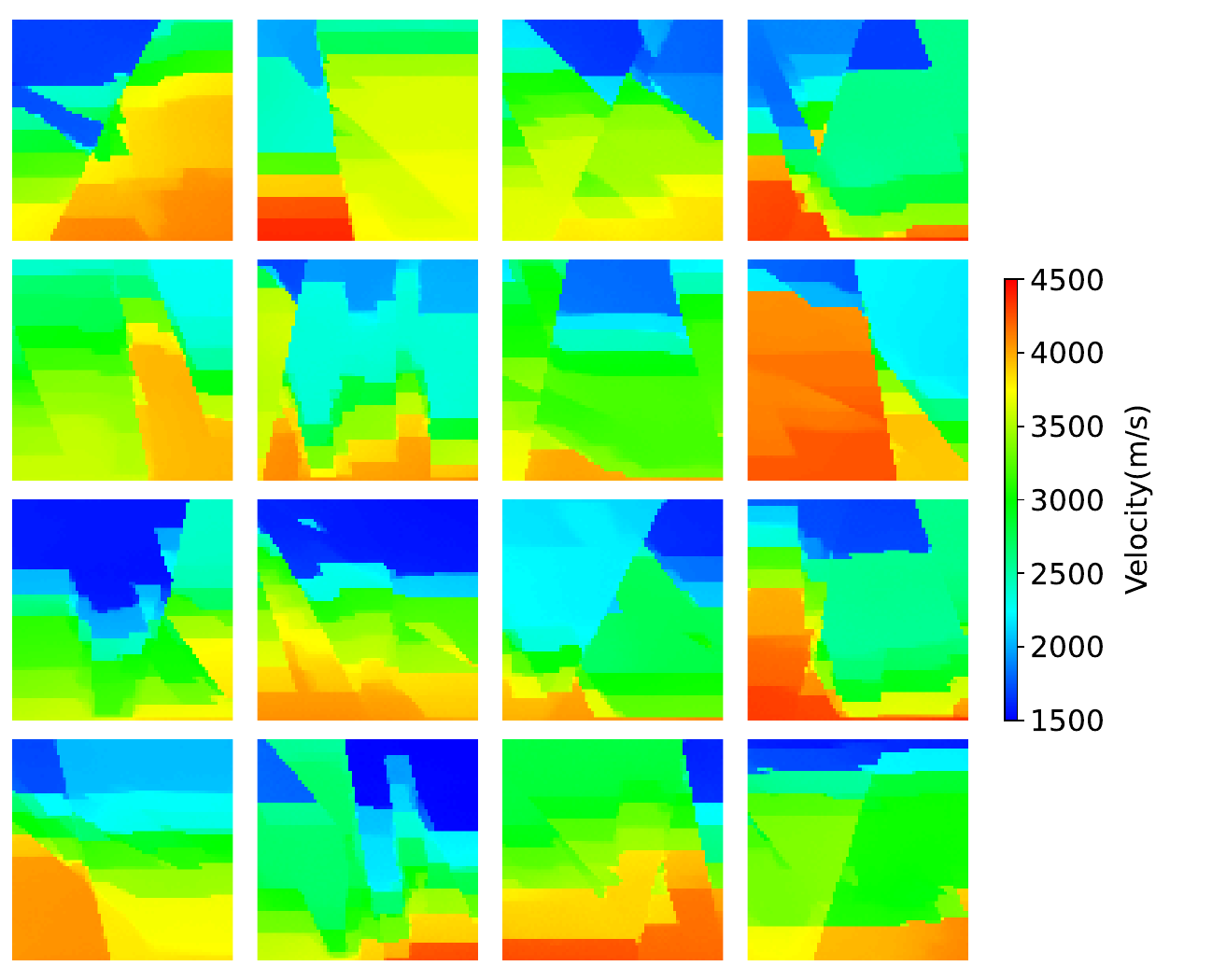}\label{fig:geo_0.5}}\quad
    \subfloat[]{\includegraphics[width=0.45\columnwidth]{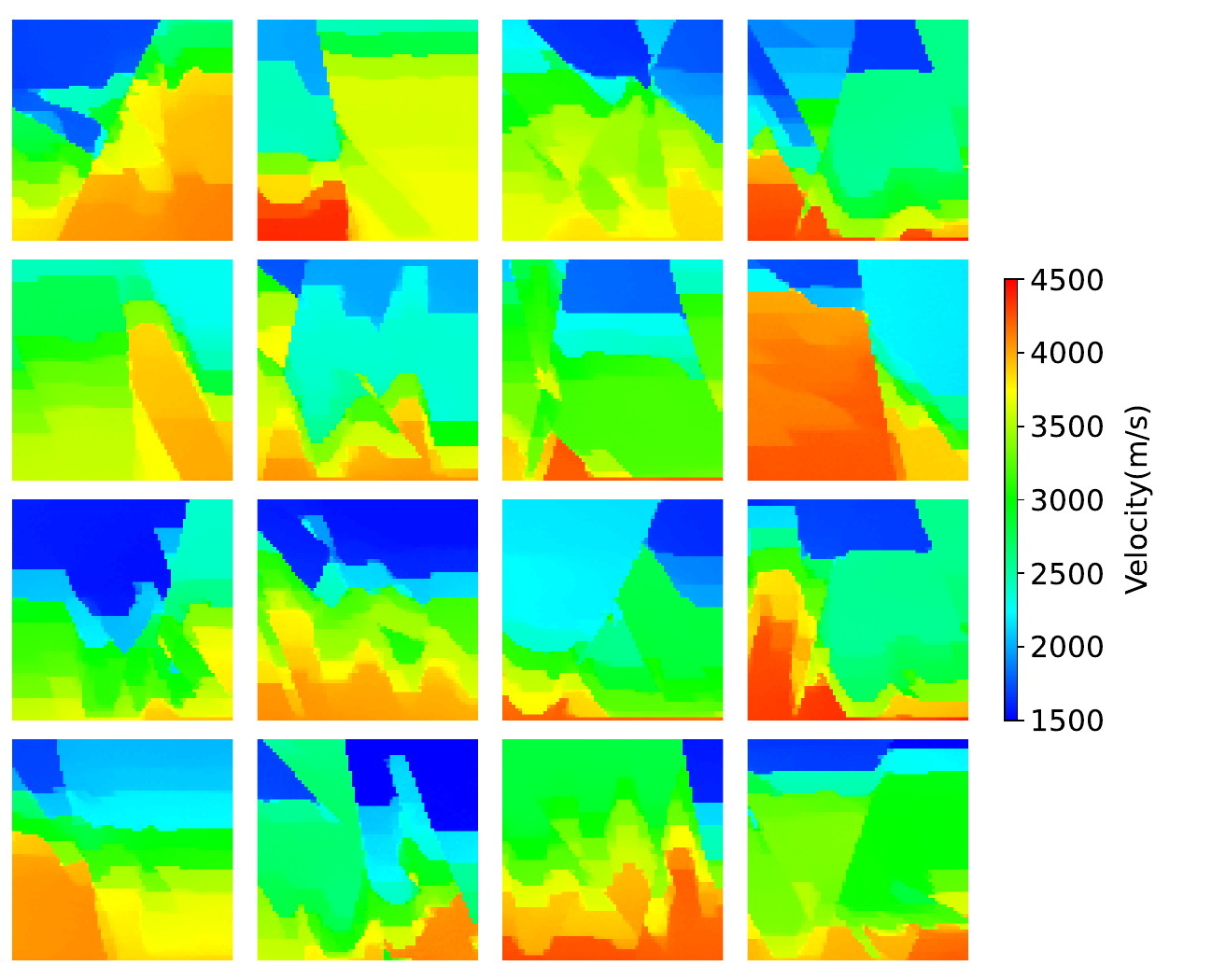}\label{fig:geo_1}}
    \caption{The generation results on the FlatFaut-B dataset by integrating the seismic-data GDM trained on the CurveFault-B dataset and the geology-oriented GDM trained on the velocity models with expected geological knowledge. (a) The ground-truth models, (b) the generated velocity models when $\lambda_1 $ = 0, (c) the generated velocity models when  $\lambda_1 $ = 0.5, and (d) the generated velocity models when   $\lambda_1 $ =1}
    \label{fig:geo_control}
\end{figure*}

We then test the impact of the geological-knowledge integration on the generation results by using the FlatFault-B dataset. Figure \ref{fig:geo_control}(a) shows the true velocity models. We use the velocity models in the FlatFault-B dataset to train the geology-oriented GDM to learn the prior geological knowledge of flat layers and faults. The trained geology-oriented GDM is then incorporated into the sampling process of the seismic-data GDM pre-trained with the CurveFault-B dataset to adapt the generation to the learned geological knowledge. The weighting factors $\lambda_1$ in Equation \ref{eq:17} is used to control their contribution.

Figure~\ref{fig:geo_control}(\subref{fig:geo_0}) shows  4×4 velocity models from the generation result when $\lambda_1=0$. We can see that the generated velocity models mainly consist of flat layers and faults which satisfy the geological knowledge of the subsurface learned by the geology-oriented GDM from the FlatFault-B velocity models. However, the accuracy of the generated model is limited in the absence of seismic data as condition. Figure~\ref{fig:geo_control}(\subref{fig:geo_1}) displays the generated velocity models when $\lambda_1$ is 1. We can see that the generated velocity model look generally close to the true velocity model with seismic data to condition the generation. However, the expected geological features of the FlatFault-B dataset are not incorporated into the generated velocity models. Figure~\ref{fig:geo_control}(\subref{fig:geo_0.5}) shows the generated velocity models when $\lambda_1$ is 0.5. The generated velocity models show consistent geological knowledge with that of the FlatFault-B models, and the velocity values are well constrained by seismic data. This indicates a reasonable integration of the seismic-data constraint and the prior geological knowledge.

\subsection{Integration of well logs information}
\begin{figure*}[htpb!]
    \centering
    \subfloat[]{\includegraphics[width=0.50\columnwidth]{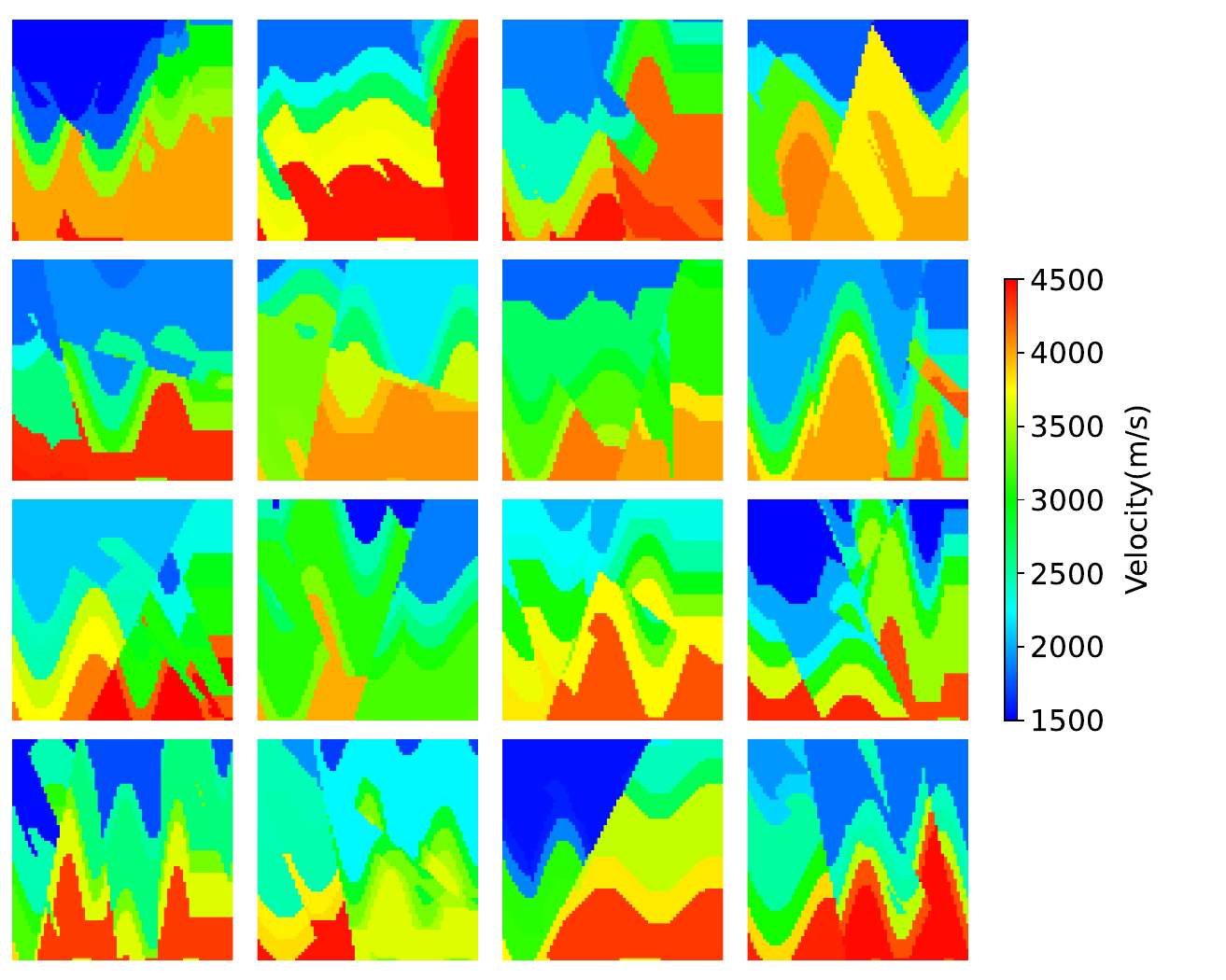}\label{fig:4logs_label}}\\
    \subfloat[]{\includegraphics[width=0.50\columnwidth]{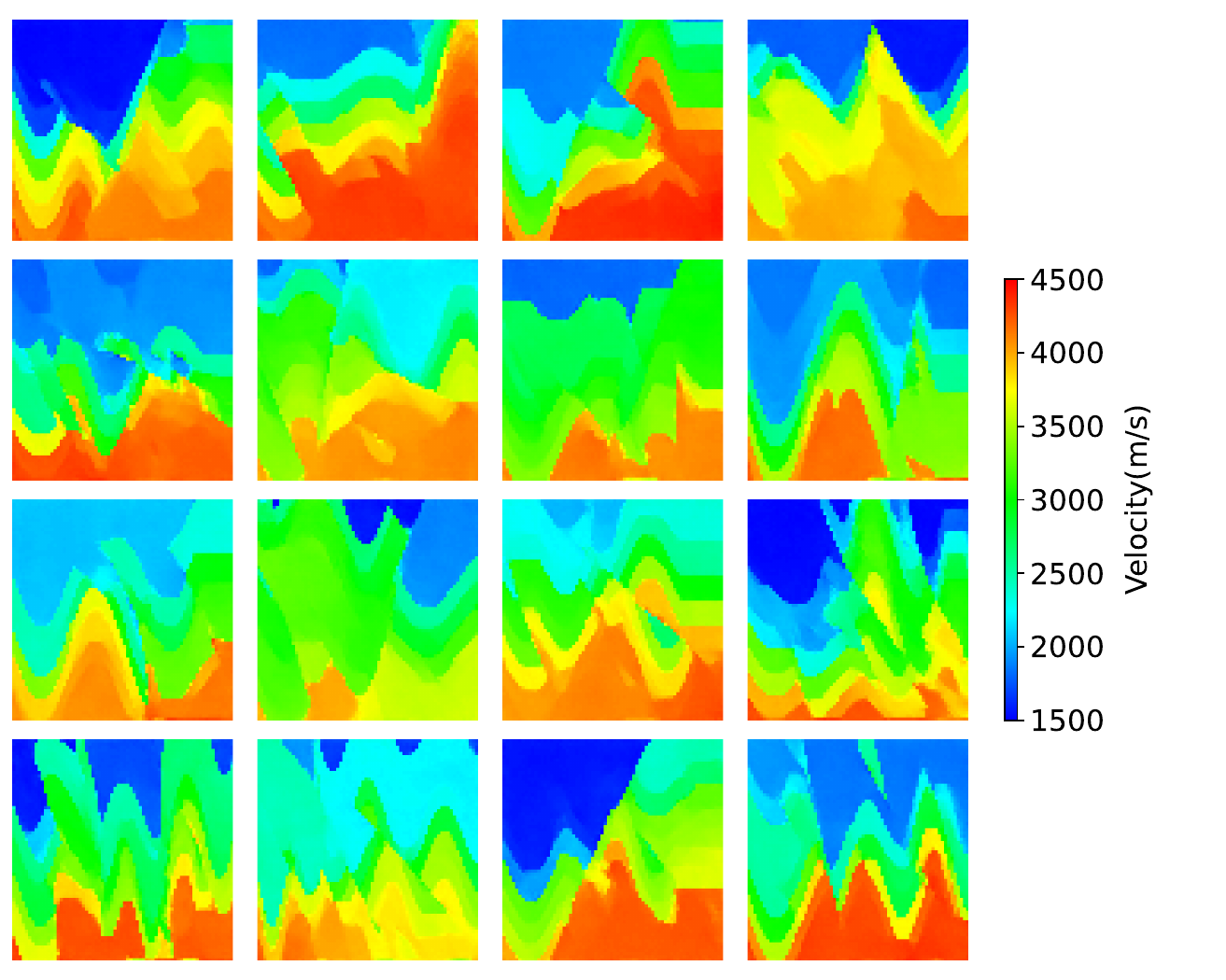}\label{fig:4logs_lowf}}
    \subfloat[]{\includegraphics[width=0.50\columnwidth]{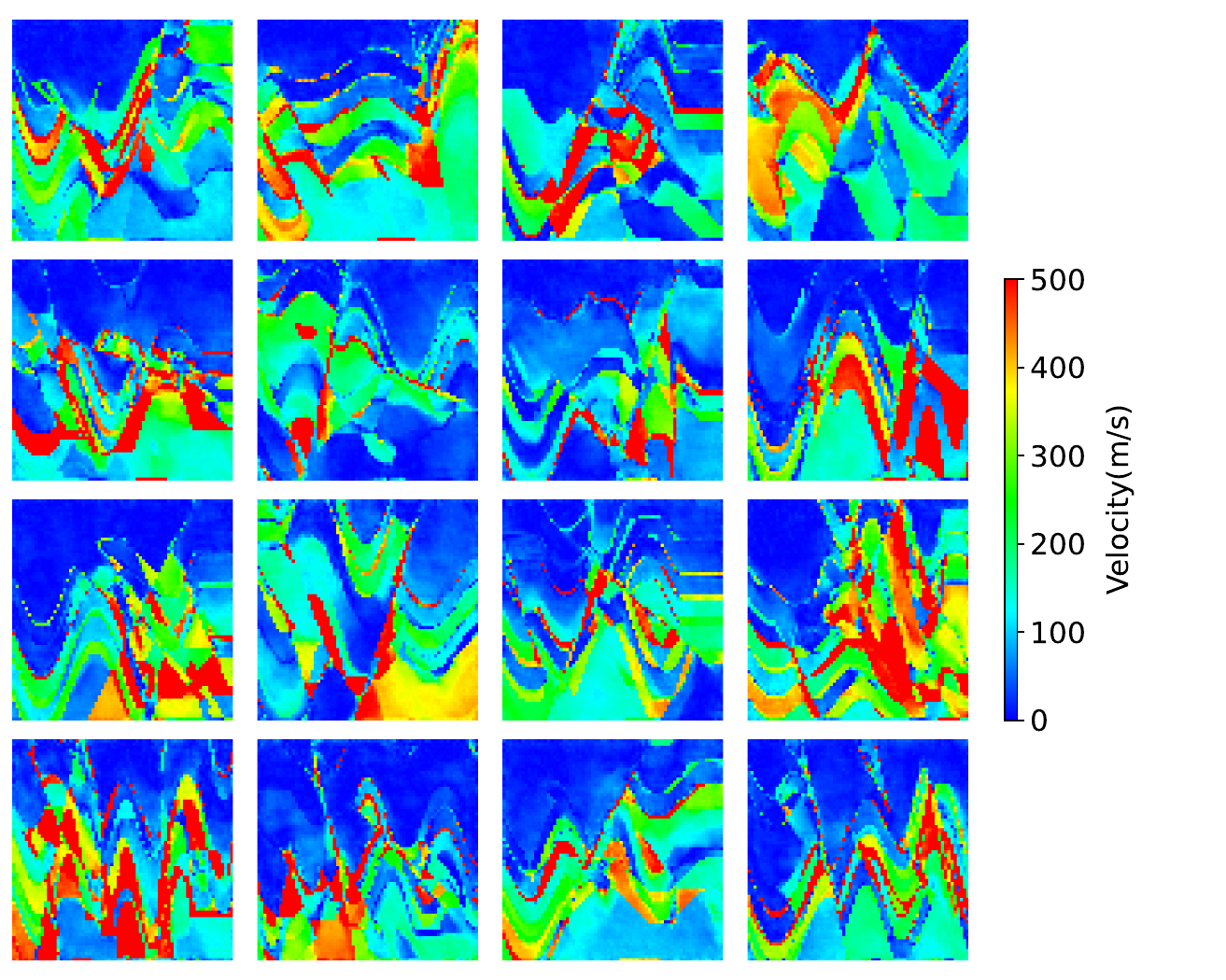}\label{fig:4logs_lowferror}}\\
    \subfloat[]{\includegraphics[width=0.50\columnwidth]{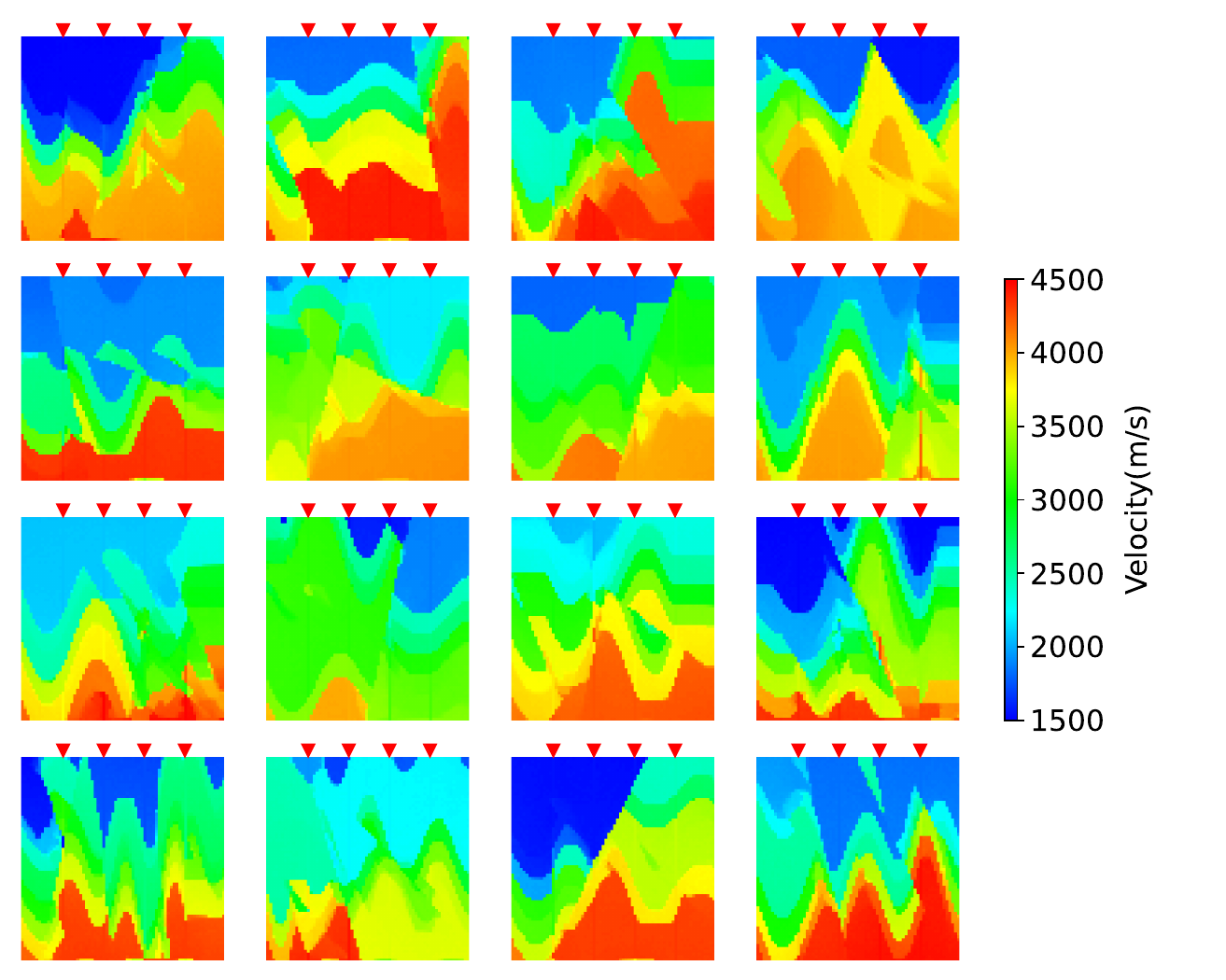}\label{fig:0.5_4logs_0}}
    \subfloat[]{\includegraphics[width=0.50\columnwidth]{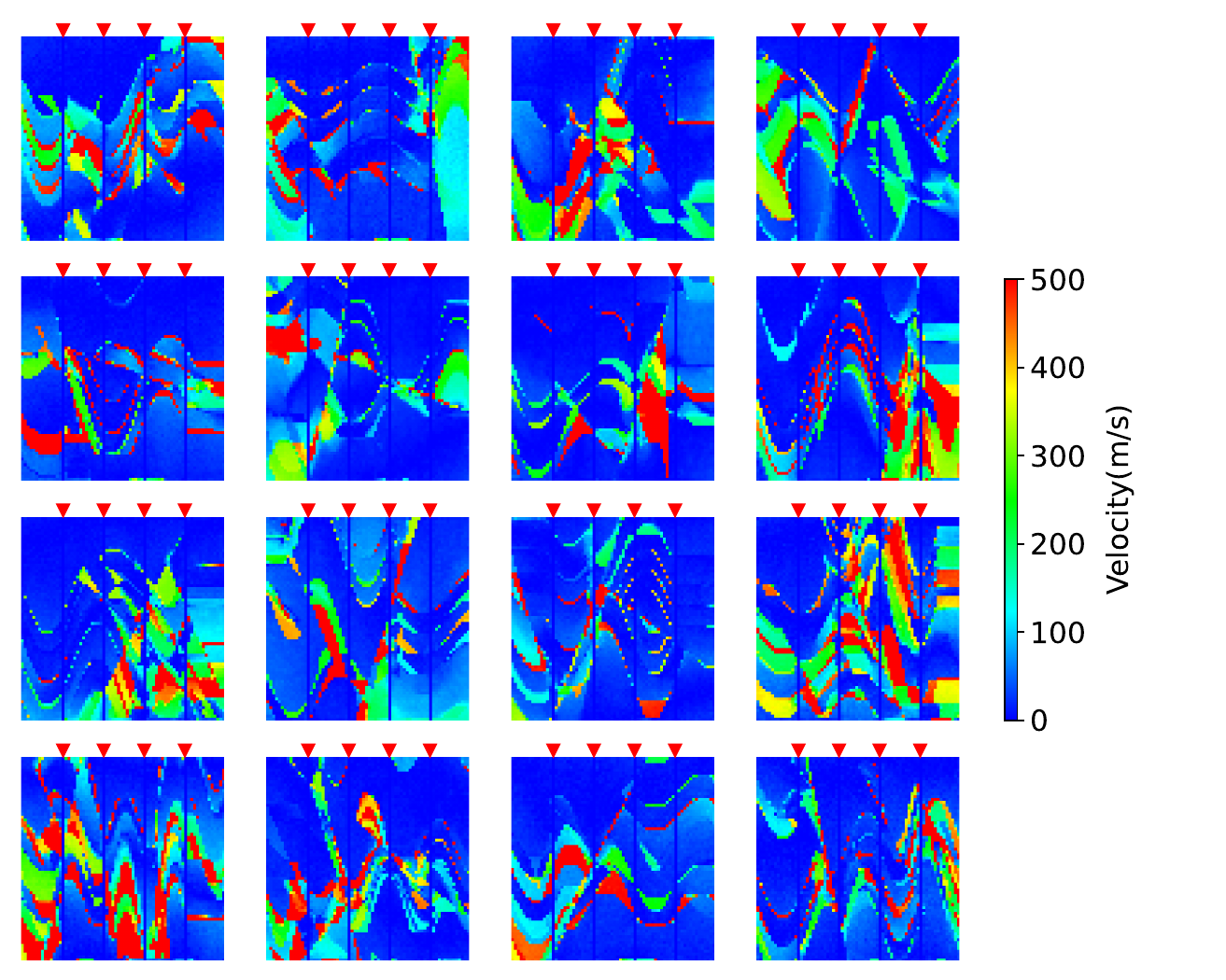}\label{fig:0.5_4logs_0error}}
    
    \caption{The generation results by integrating the seismic-data GDM and the well-log GDM. (a) The ground-truth models, (b) the generated velocity models when $\lambda_1 $ = 1, (c) residuals between (b) and (a), (d) the generated velocity models when $\lambda_1 $ = 0.5, and (e) residuals between (d) and (a).}
    \label{fig:well_logs_coefficient}
\end{figure*}
\begin{figure*}[htpb!]
   \centering
    \subfloat[]{\includegraphics[width=0.3\columnwidth]{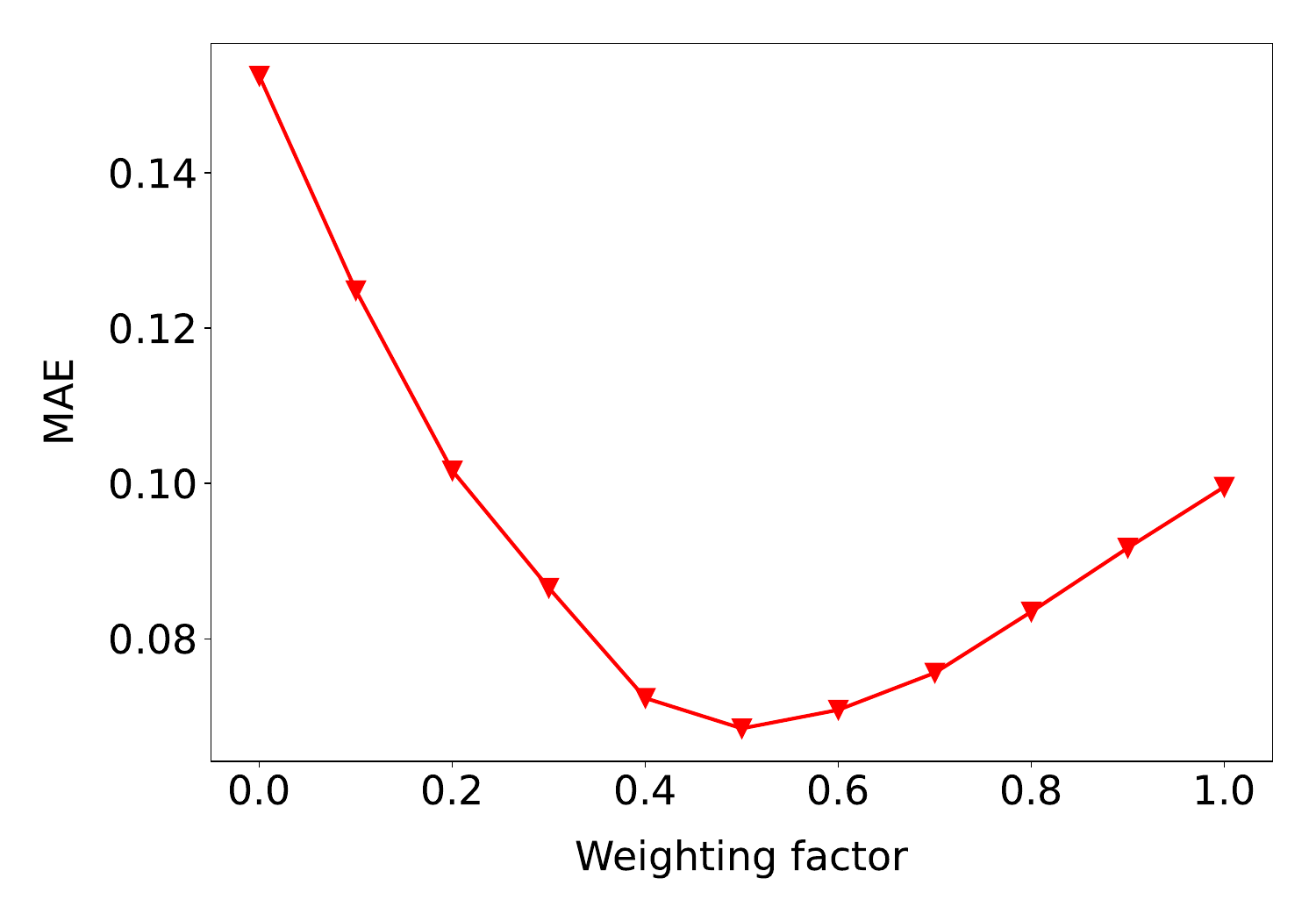}\label{0.8_4logs_label}}
      \subfloat[]{\includegraphics[width=0.3\columnwidth]{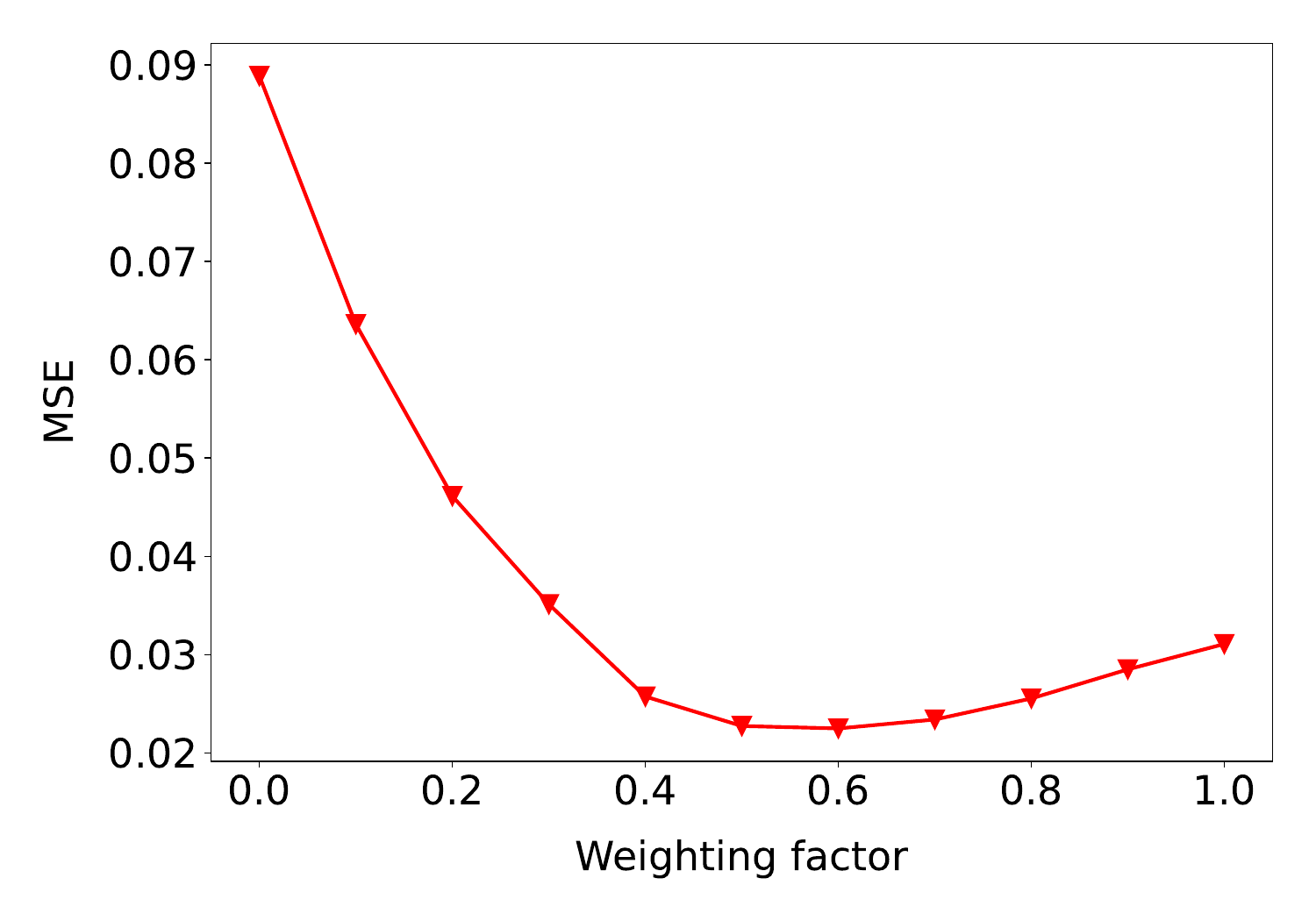}\label{0.8_4logs_label}}
        \subfloat[]{\includegraphics[width=0.3\columnwidth]{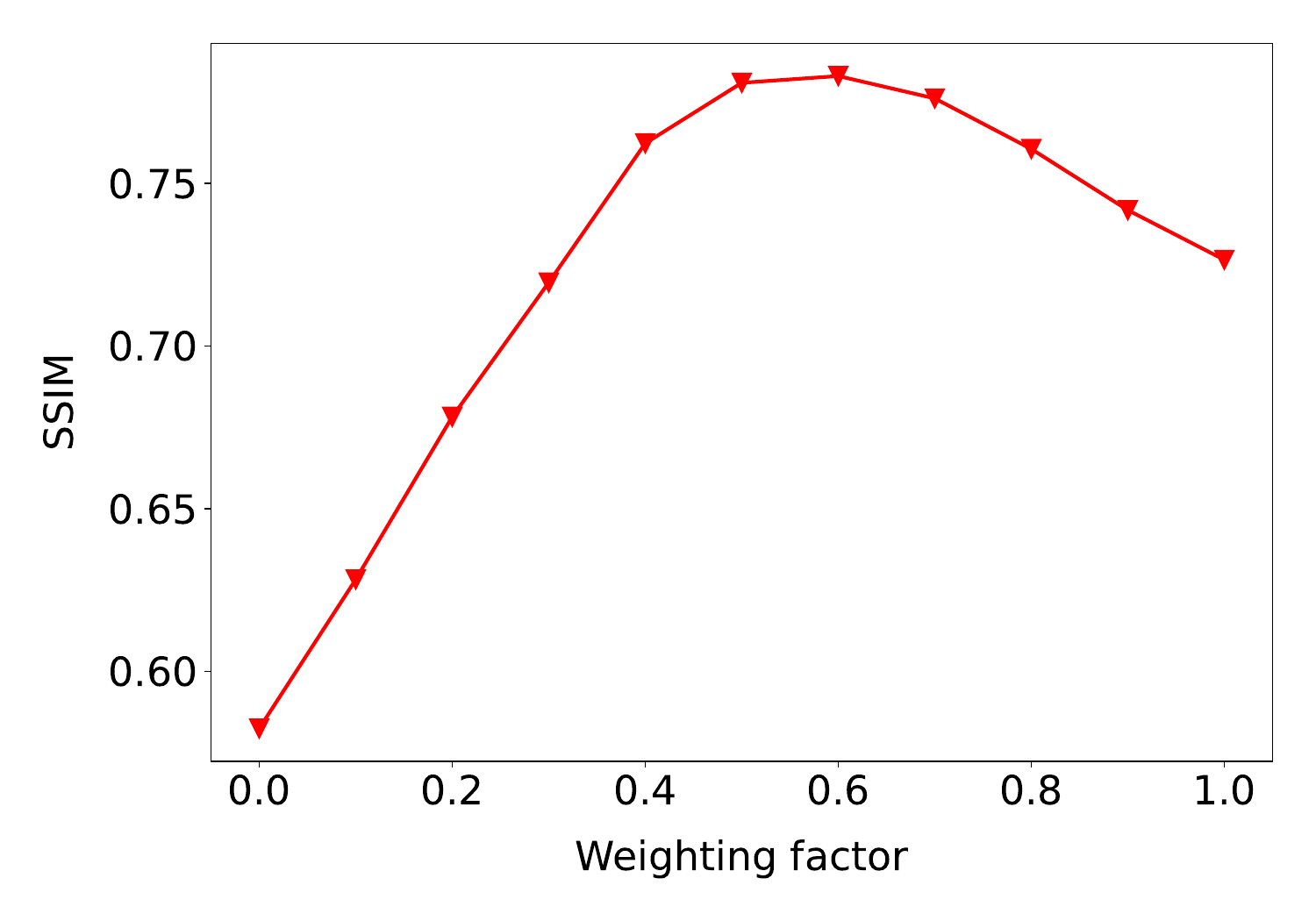}\label{0.8_4logs_label}}
    \caption{Quantitative comparison of the generation results with different weighting factors $ \lambda_ 2$ for the well-log integration. (a) MAE, (b) MSE, and (c) SSIM.}
    \label{fig:well_logs_metrics}
\end{figure*}

Here, we assume that several well logs are available. To adapt the generated model to the available well information, we train a well-log GDM by using the CurveFault-B velocity models as the target and the well logs as the condition. The well-log GDM is then integrated into the sampling process of the seismic-data GDM to enable its generated model to respect the available well logs. Figure~\ref{fig:well_logs_coefficient}(\subref{fig:4logs_label}) shows the  4×4 velocity models from the CurveFault-B dataset. We first select 4 traces evenly from the true model as the available well logs. We compute the quality metrics with different factors between 0 and 1 to test how the weighting factor $\lambda_2$ in Equation \ref{eq:18} influence the integrated result. The quality metric curves are shown in Figure~\ref{fig:well_logs_metrics}. We can see that the generated model tend to be optimal when the weighting factor is close to 0.5.

Figure~\ref{fig:well_logs_coefficient}(\subref{fig:4logs_lowf}) and (\subref{fig:4logs_lowferror}) show the generated velocity models when $\lambda_{2}=1$ and the residuals between these generated model and the true models, respectively. We can see that the generative model without well-logs integration fails to estimate the subsurface velocity accurately, especially for the deep complex structure. Figure~\ref{fig:well_logs_coefficient}(\subref{fig:0.5_4logs_0}) and (\subref{fig:0.5_4logs_0error}) show the generated velocity models when $\lambda_{2}=0.5$ and their velocity residuals, respectively. For an intuitive observance of the integration, we incorporate the available well logs into the generated velocity models. We can see that the generated velocity models are well adapted to the constraints imposed by the well-log information. The velocity models are generated with improved accuracy.

\subsection{Hess VTI model example}
\begin{figure*}[htpb!]
  \centering
   \subfloat[]{\includegraphics[width=0.28\columnwidth]{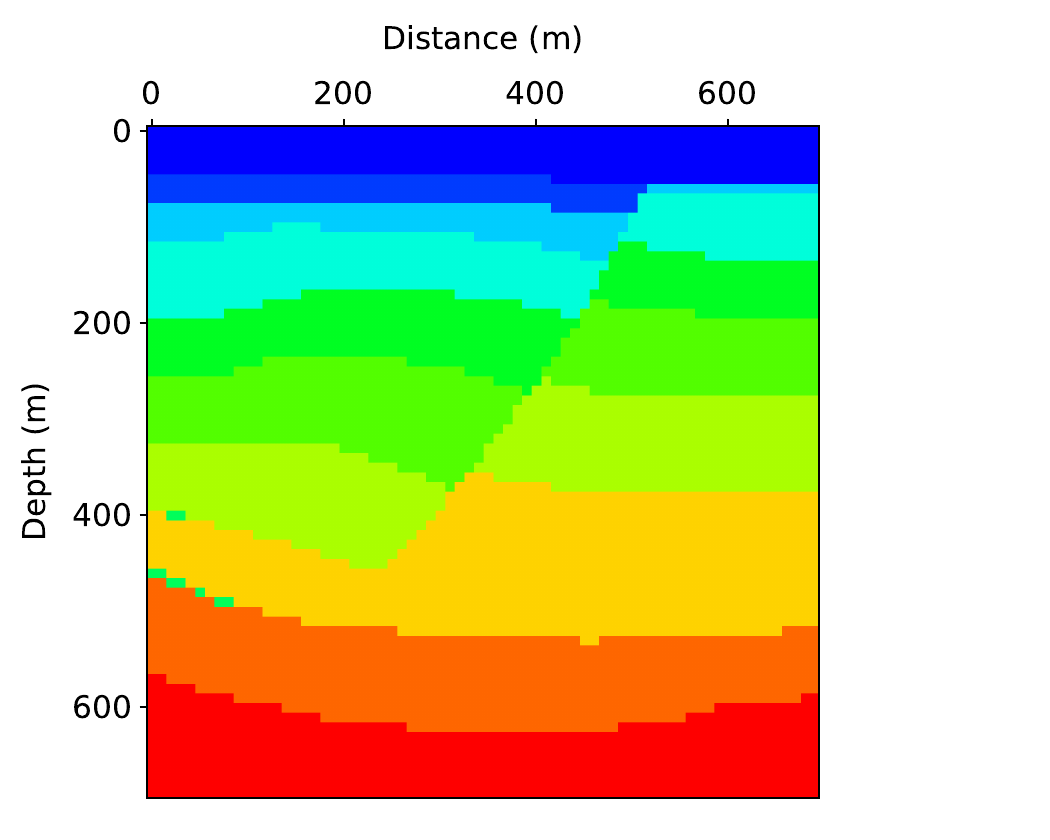}\label{fig:timodel_label}}
    \hspace{-30pt}
    \subfloat[]{\includegraphics[width=0.28\columnwidth]{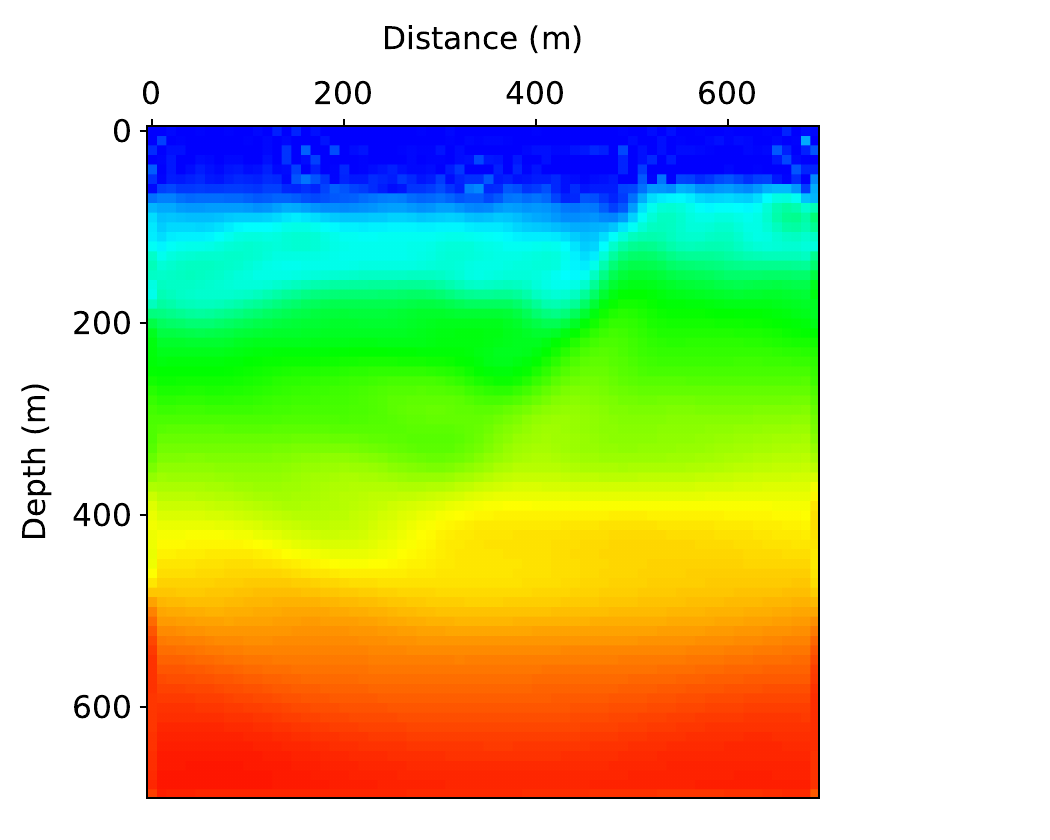}\label{fig:timodel_FWI}}
    \hspace{-30pt}
    \subfloat[]{\includegraphics[width=0.28\columnwidth]{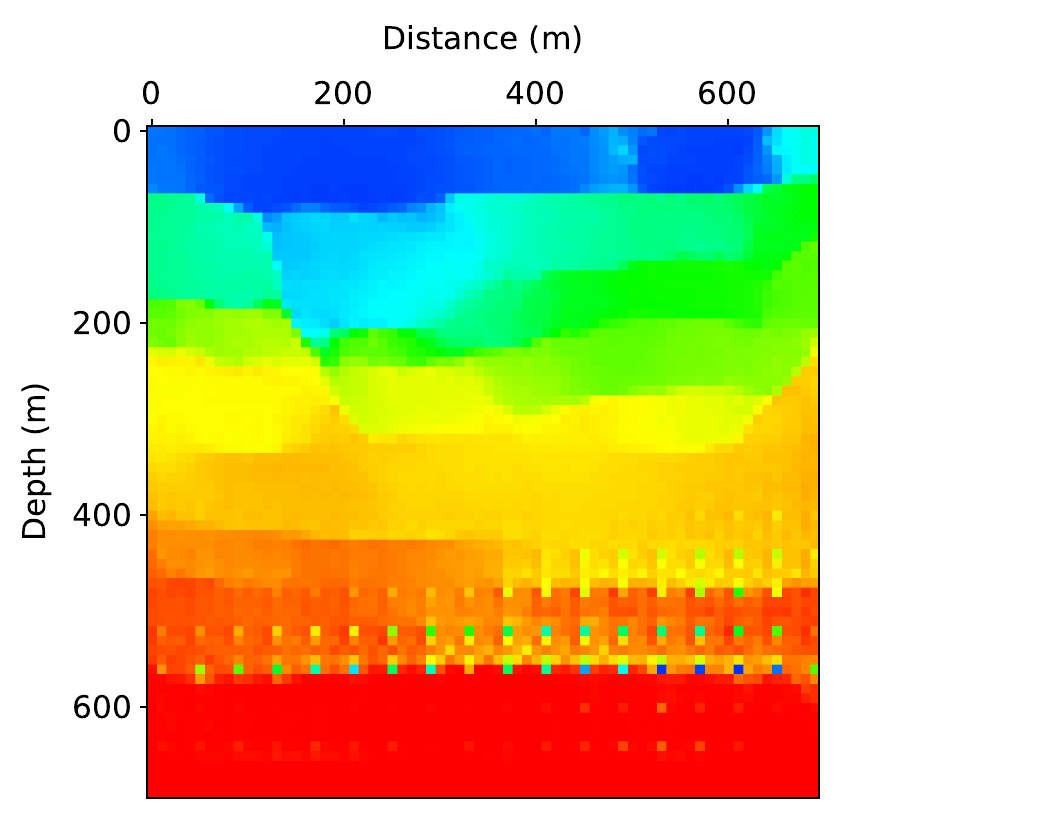}\label{fig:timodel_InversionNet}}
    \hspace{-30pt}
    \subfloat[]{\includegraphics[width=0.28\columnwidth]{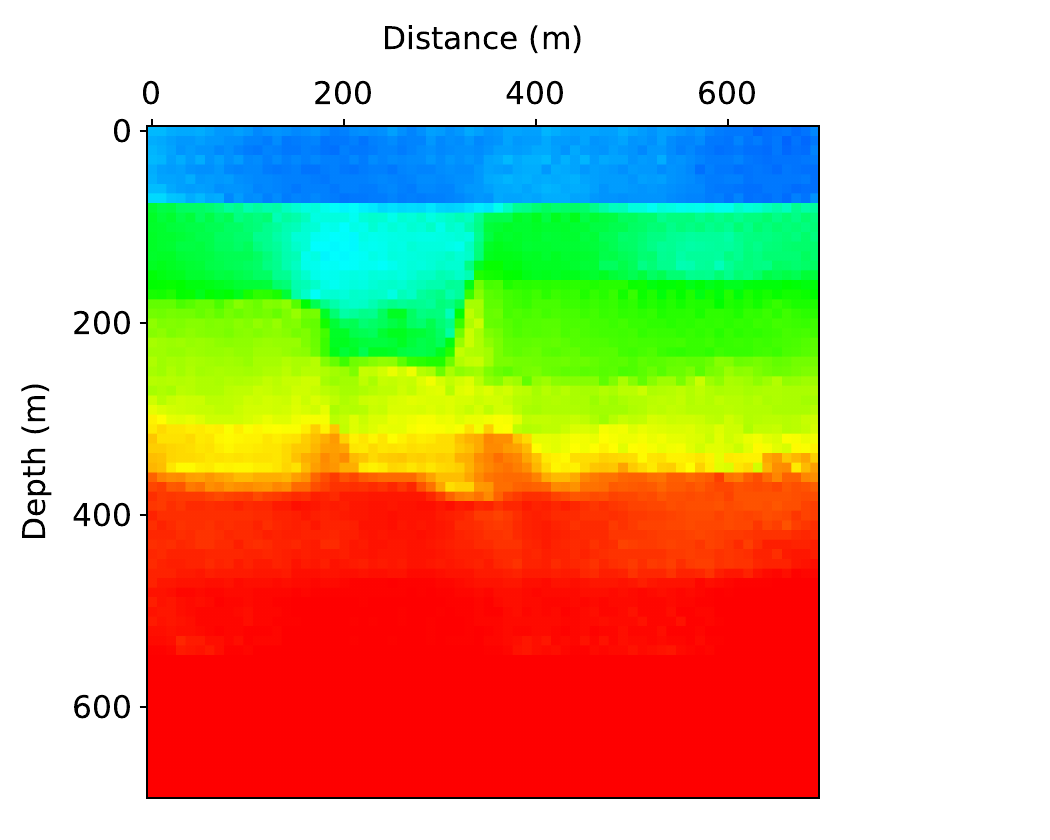}\label{fig:timodel_VelocityGAN}}\\
    \subfloat[]{\includegraphics[width=0.28\columnwidth]{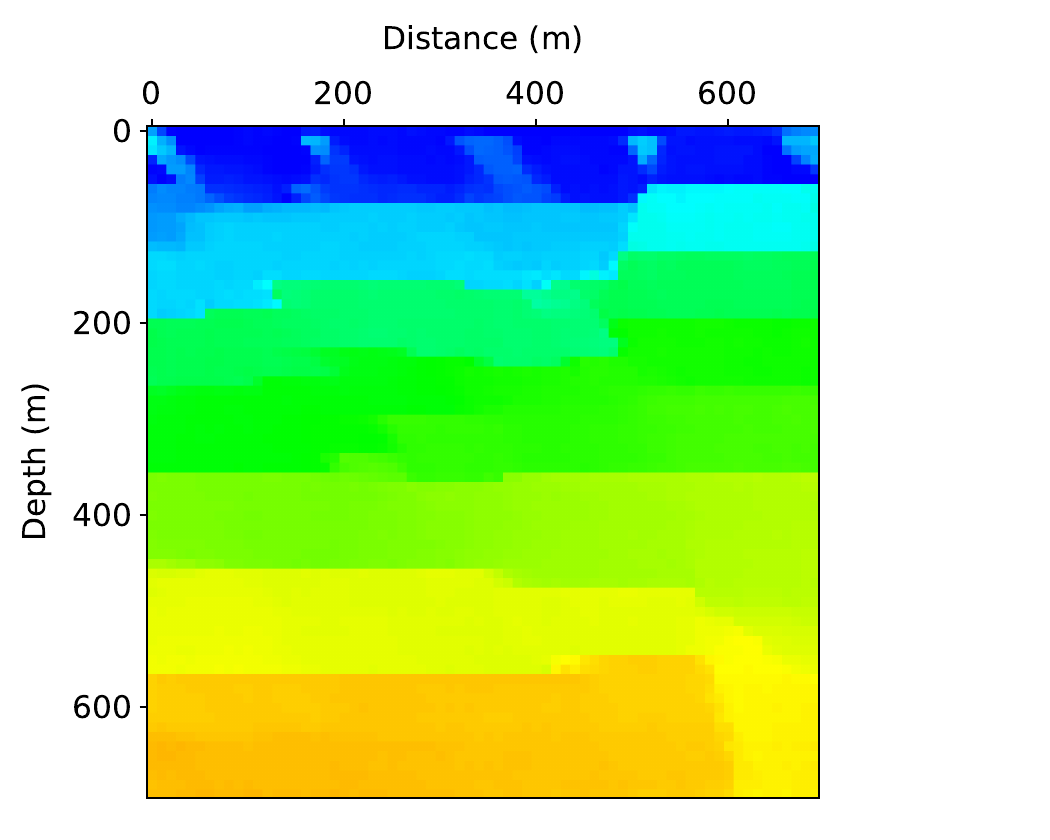}\label{fig:timodel_original}}
       \hspace{-30pt}
    \subfloat[]{\includegraphics[width=0.28\columnwidth]{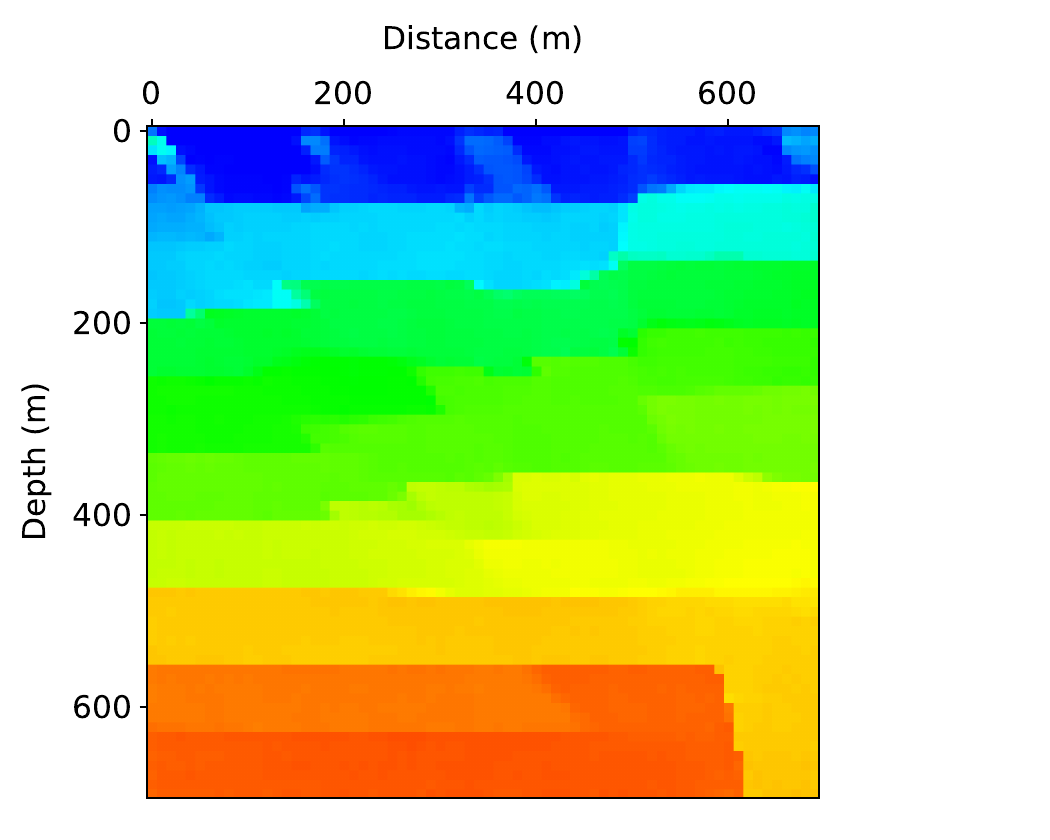}\label{fig:timodel_back_int}}
       \hspace{-30pt}
    \subfloat[]{\includegraphics[width=0.28\columnwidth]{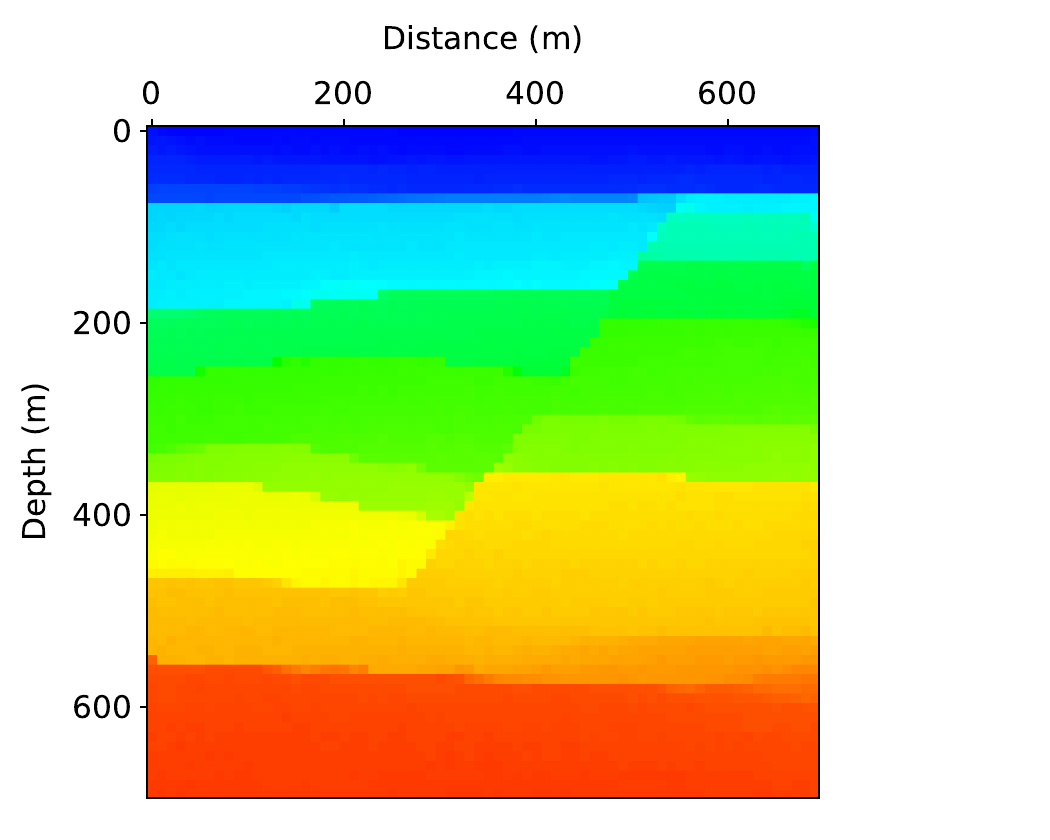}\label{fig:timodel_geo_int}}
       \hspace{-30pt}
    \subfloat[]{\includegraphics[width=0.28\columnwidth]{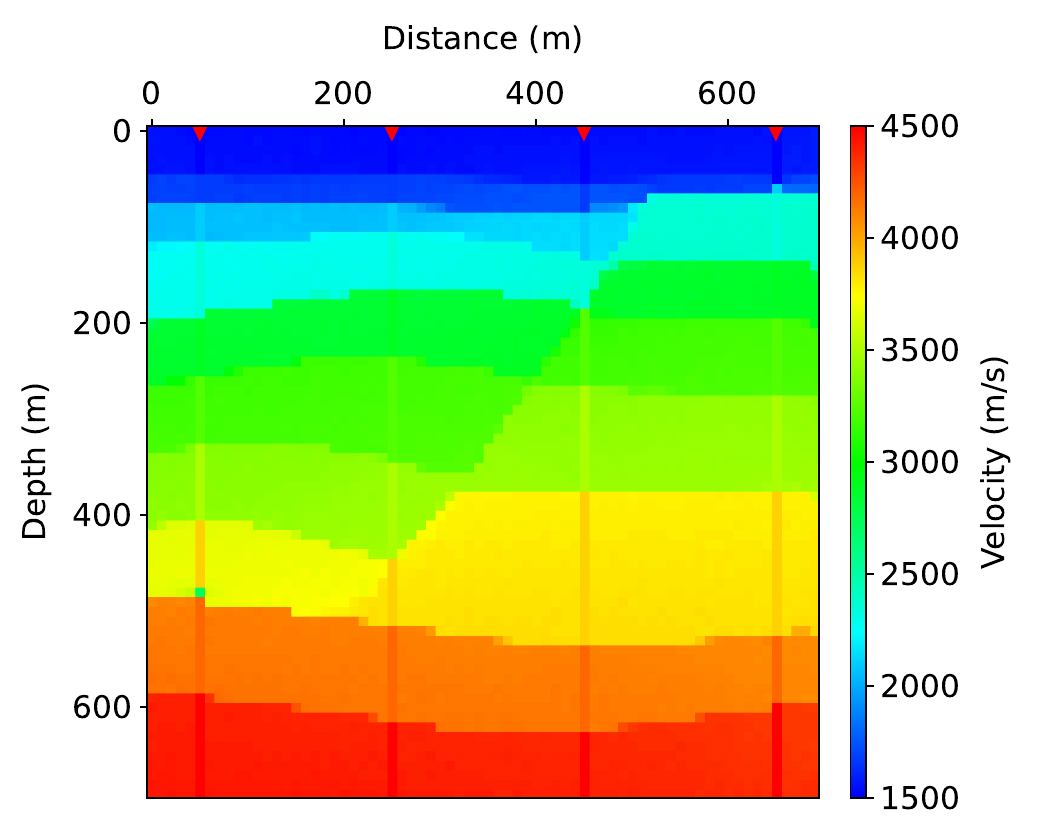}\label{fig:timodel_log_int}}    
    \caption{Inversion results on the Hess VTI model example using different methods. (a) The true velocity model, (b)  conventional FWI, (c) InversionNet, (d) VelocityGAN, (e) the seismic-data GDM, (f) the seismic-data GDM with background velocity, (g) integration of the seismic-data GDM, background velocity, and the geology-oriented GDM when the weighting factor $\lambda_1=0.5$, and (h) integration of the seismic-data GDM, background velocity, and the well-log GDM when the weighting factor $\lambda_2=0.5$}
    \label{fig:timodel_all}
    
\end{figure*}

Finally, we apply the proposed approach on the Hess VTI model example to validate its generalization. We extract a part of the P-velocity model and resample the model size to that of the OpenFWI datasets. As shown in Figure~\ref{fig:timodel_all}(\subref{fig:timodel_label}), the extracted velocity model consists of multiple layers and a fault oriented towards left. The forward modeling technique with same parameter setting for simulating the OpenFWI datasets is also applied to synthesize the observed seismic data. We use the FlatFault-B dataset to train the seismic-data GDM to learn the velocity model distribution under the seismic data as condition. We then use the observed seismic data to condition the seismic-data GDM to generate the velocity model. Figure~\ref{fig:timodel_all}(\subref{fig:timodel_original}) shows the generated velocity model. We can see that the velocity model is generated with flat layers and a short fault. However, the seismic-data GDM fails to reconstruct the curved layers because the training dataset do not involve any curved layers. Besides, the velocities of the deep layers are not estimated accurately.
\begin{figure*} [htbp!]
	\centering
		\includegraphics[width=0.7\columnwidth]{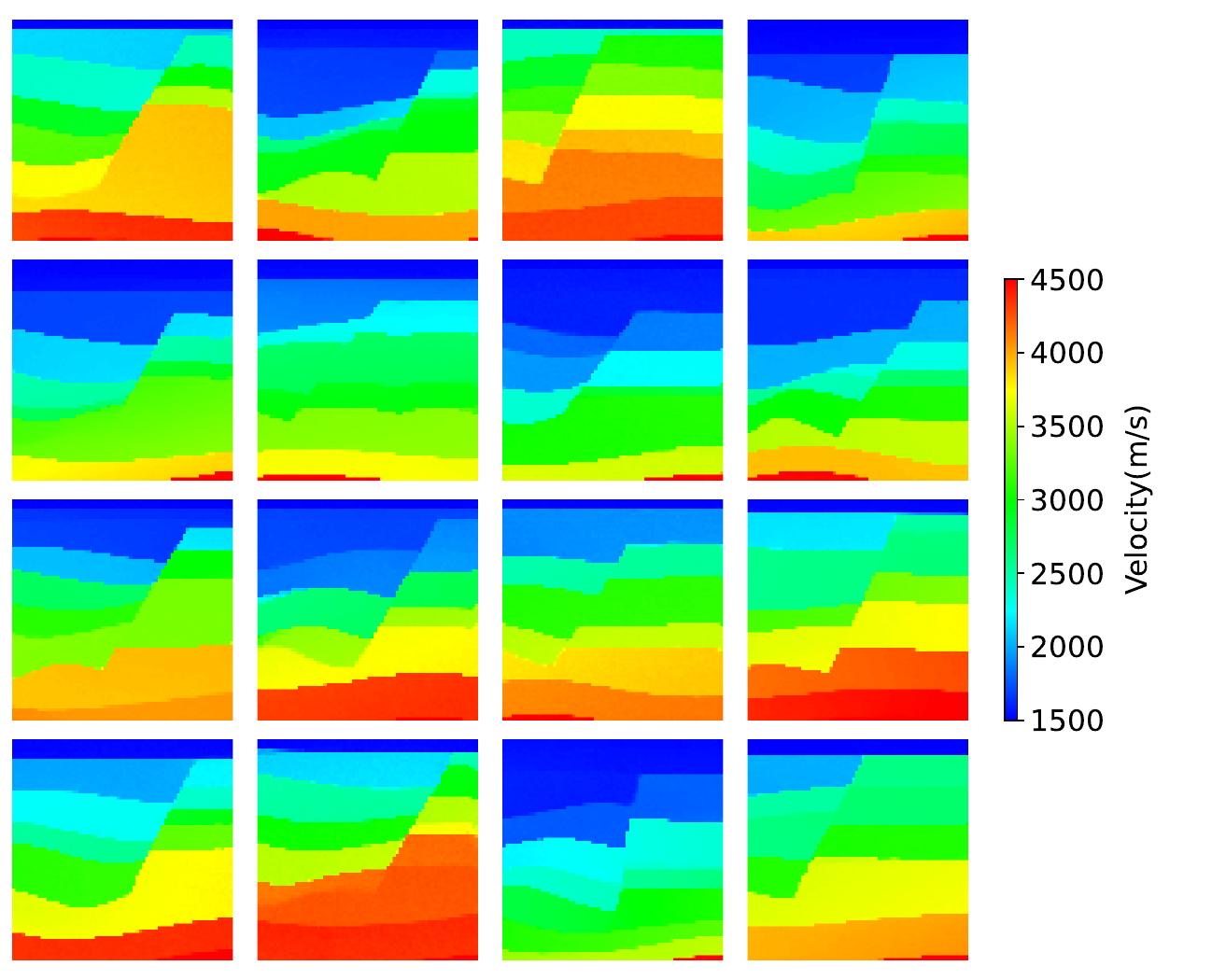}
\caption{Generation examples of the geology-oriented GDM}
\label{fig:timodel_unconditional}
\end{figure*}

\begin{figure*} [htpb!]
	\centering
	
 	\includegraphics[width=0.5\columnwidth]{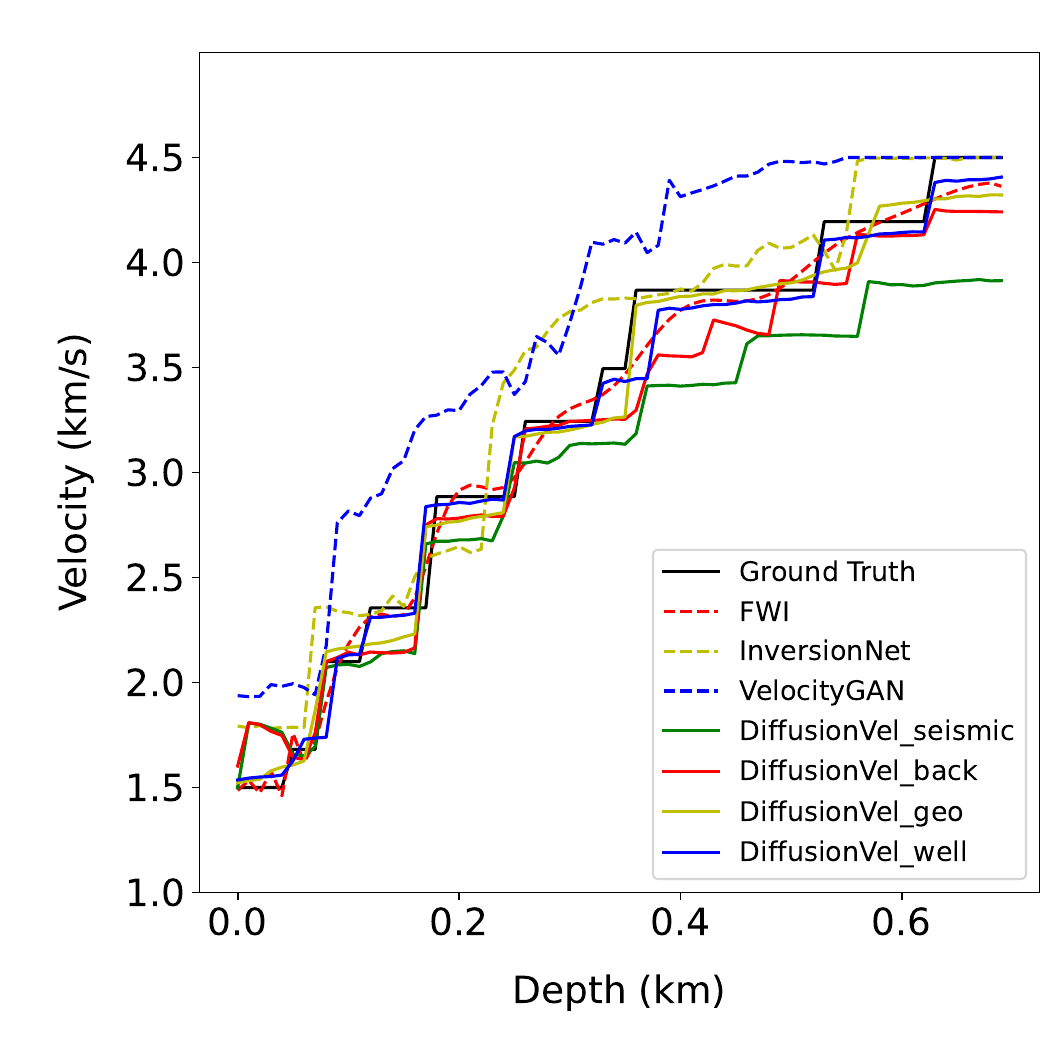}
          
\caption{The vertical velocity profiles at the center of generated models in Figure~\ref{fig:timodel_all}.}
 \label{fig:timodel:single_trace}
\end{figure*}

We smooth the true velocity model by using a Gaussian filter with a kernel size of 29 to obtain the background model. We then incorporate the background velocity model into the first 8 sampling steps of the seismic-data GDM. Figure~\ref{fig:timodel_all}(\subref{fig:timodel_back_int}) shows the generated velocity model with the integration of the background model. We can see that the background model helps to improve the generation accuracy, especially in the deep area. Subsequently, to integrate the prior geological information, we prepare a velocity model distribution (2000 velocity models) sharing the following geological features: (1) the velocity increases with depth, (2) the layers are slightly curved, and (3) one fault oriented towards the left and terminating before the final layers. These 2000 velocity models are used to train the geology-oriented GDM to learn the prior geological knowledge. Figure~\ref{fig:timodel_unconditional} shows the generation example of this geology-oriented GDM. In addition to the background model integration, we integrate the geology-oriented GDM into the sampling process of the seismic-data GDM with the weighting factor $\lambda_1$ of 0.5. Figure~\ref{fig:timodel_all}(\subref{fig:timodel_geo_int}) shows the velocity model generated by integrating the geology-oriented GDM into the seismic-data GDM. We can see that the fault and the curved layers are well recovered in the generated velocity model, showing consistent subsurface features with the true velocity model. However, the layer interfaces are not positioned accurately. We thus integrate the available well logs into the generation. Assuming a few well logs are available, we use the prepared dataset previously used in the geology-oriented GDM to train a well-log GDM. The well-log GDM is then integrated into the sampling process of the seismic-data GDM with the weighting factor $\lambda_2$ of 0.5. Four traces from the true velocity model at the 50 m, 250 m, 450 m, and 650 m distance are selected as the available well logs. Figure~\ref{fig:timodel_all}(\subref{fig:timodel_log_int}) shows the final generated velocity model with the integration of the seismic-data GDM, the background velocity model, and the well-log GDM. We can see that the generated velocity model looks close to the true velocity model.

Then, we make a comparison between our proposed method and the conventional FWI, InversionNet and VelocityGAN by using the same FlatFault-B dataset as training dataset. Starting from the background model, the conventional FWI recover the velocity model shown in Figure~\ref{fig:timodel_all}(\subref{fig:timodel_FWI}) with relatively-low accuracy and resolution because of the limited seismic data. Figure~\ref{fig:timodel_all}(\subref{fig:timodel_InversionNet}) and (\subref{fig:timodel_VelocityGAN}) show the predicted velocity models by using InversionNet and VelocityGAN, respectively. We can see that the important structures including a fault and multiple layers are not recovered successfully, especially for the deep part. For detailed comparison, we also show the vertical velocity profiles at 350 m distance of these generated models in Figure~\ref{fig:timodel:single_trace}.

\section{Discussion}

We propose a data-driven velocity inversion method, DiffusionVel, by effectively integrating multiple information including seismic data, background velocity, geological knowledge, and well information into the sampling process of the GDM. We use the seismic-data GDM and the well-log GDM to adapt the generated velocity model to their respective constraint. To take advantage of the known background model, we apply a low-pass filter to obtain the LF components of the generated velocity model and then replace it with the given background velocity model. We use the geology-oriented GDM to constrain the generated velocity model with prior geological knowledge. In our DiffusionVel method, the multiple information is integrated into the sampling process of GDM. At each sampling step, the generated velocity models of the GDMs in DiffusionVel are combined together through a weighted summation. Thus, we can flexibly control the constraints of each information by adjusting the weighting factors. The final velocity model of multi-information integration is generated at the last sampling step. The test results on the OpenFWI datasetS and the Hess VTI example demonstrate that our DiffusionVel method allow us to integrate the multiple information about the subsurface and predict the velocity model reasonably. 

When integrating prior geological knowledge and well logs, we select a fixed weighting factor throughout the sampling process. To achieve better integrated generation, we can adopt a weighting schedule where the weighting factor varies with the sampling step. The weighting schedule allows us to control the constraint from each information more flexibly.

The key to integrate the background velocity is a reasonable selection of the low-pass filter. In the numerical examples, a known low-pass filter is applied to the true model to obtain the background model. In the real cases, we need to design the low-pass filter based on trial and error to produce the best integration results. The experiment on geological-knowledge integration demonstrates that we can incorporate a geology-oriented GDM that learns the prior geological knowledge to improve the generalization of the seismic-data GDM. Furthermore, we can improve the quality and size of the training dataset to achieve better generalization.

\section{Conclusion}

We develop DiffusionVel, a data-driven velocity inversion approach based on the GDMs with integration of multiple information including seismic data, background velocity, geological knowledge, and well logs. Test results on the OpenFWI datasets and the Hess VTI model example demonstrate that DiffusionVel generates the velocity model with higher accuracy than the conventional FWI, InversionNet and VelocityGAN. The multi-information integration further improves the generalization of the data-driven velocity inversion approach.

\bibliographystyle{unsrt}  
\bibliography{example}

\newcommand{\SortNoop}[1]{}
\begin{thebibliography}{10}

\bibitem{tarantola1984inversion}
Albert Tarantola.
\newblock Inversion of seismic reflection data in the acoustic approximation.
\newblock {\em Geophysics}, 49(8):1259--1266, 1984.

\bibitem{virieux2009overview}
Jean Virieux and St{\'e}phane Operto.
\newblock An overview of full-waveform inversion in exploration geophysics.
\newblock {\em Geophysics}, 74(6):WCC1--WCC26, 2009.

\bibitem{golub1999tikhonov}
Gene~H Golub, Per~Christian Hansen, and Dianne~P O'Leary.
\newblock Tikhonov regularization and total least squares.
\newblock {\em SIAM journal on matrix analysis and applications}, 21(1):185--194, 1999.

\bibitem{strong2003edge}
David Strong and Tony Chan.
\newblock Edge-preserving and scale-dependent properties of total variation regularization.
\newblock {\em Inverse problems}, 19(6):S165, 2003.

\bibitem{asnaashari2013regularized}
Amir Asnaashari, Romain Brossier, St{\'e}phane Garambois, Fran{\c{c}}ois Audebert, Pierre Thore, and Jean Virieux.
\newblock Regularized seismic full waveform inversion with prior model information.
\newblock {\em Geophysics}, 78(2):R25--R36, 2013.

\bibitem{zhang2022regularized}
Zhendong Zhang and Tariq Alkhalifah.
\newblock Regularized elastic full-waveform inversion using deep learning.
\newblock In {\em Advances in subsurface data analytics}, pages 219--250. Elsevier, 2022.

\bibitem{li2016integrated}
Yunyue Li, Biondo Biondi, Robert Clapp, and Dave Nichols.
\newblock Integrated vti model building with seismic data, geologic information, and rock-physics modeling—part 1: Theory and synthetic test.
\newblock {\em Geophysics}, 81(5):C177--C191, 2016.

\bibitem{li2021deep}
Yuanyuan Li, Tariq Alkhalifah, and Zhendong Zhang.
\newblock Deep-learning assisted regularized elastic full waveform inversion using the velocity distribution information from wells.
\newblock {\em Geophysical Journal International}, 226(2):1322--1335, 2021.

\bibitem{ronneberger2015u}
Olaf Ronneberger, Philipp Fischer, and Thomas Brox.
\newblock U-net: Convolutional networks for biomedical image segmentation.
\newblock In {\em Medical image computing and computer-assisted intervention--MICCAI 2015: 18th international conference, Munich, Germany, October 5-9, 2015, proceedings, part III 18}, pages 234--241. Springer, 2015.

\bibitem{edinburgh31marchenko}
Kees~Wapenaar Edinburgh and Joost van~der Neut.
\newblock Marchenko redatuming: Advantages and limitations in complex media.
\newblock In {\em Proceedings of the 15th International Congress of the Brazilian Geophysical Society \& EXPOGEF, Rio de Janeiro, Brazil}, volume~31, 2017.

\bibitem{vaswani2017attention}
Ashish Vaswani, Noam Shazeer, Niki Parmar, Jakob Uszkoreit, Llion Jones, Aidan~N Gomez, {\L}ukasz Kaiser, and Illia Polosukhin.
\newblock Attention is all you need.
\newblock {\em Advances in neural information processing systems}, 30, 2017.

\bibitem{szegedy2017inception}
Christian Szegedy, Sergey Ioffe, Vincent Vanhoucke, and Alexander Alemi.
\newblock Inception-v4, inception-resnet and the impact of residual connections on learning.
\newblock In {\em Proceedings of the AAAI conference on artificial intelligence}, volume~31, 2017.

\bibitem{lee2020maskgan}
Cheng-Han Lee, Ziwei Liu, Lingyun Wu, and Ping Luo.
\newblock Maskgan: Towards diverse and interactive facial image manipulation.
\newblock In {\em Proceedings of the IEEE/CVF conference on computer vision and pattern recognition}, pages 5549--5558, 2020.

\bibitem{he2022masked}
Kaiming He, Xinlei Chen, Saining Xie, Yanghao Li, Piotr Doll{\'a}r, and Ross Girshick.
\newblock Masked autoencoders are scalable vision learners.
\newblock In {\em Proceedings of the IEEE/CVF conference on computer vision and pattern recognition}, pages 16000--16009, 2022.

\bibitem{rasht2022}
Majid Rasht-Behesht, Christian Huber, Khemraj Shukla, and George~Em Karniadakis.
\newblock Physics-informed neural networks (pinns) for wave propagation and full waveform inversions.
\newblock {\em Journal of Geophysical Research: Solid Earth}, 127(5):e2021JB023120, 2022.
\newblock e2021JB023120 2021JB023120.

\bibitem{ovcharenko2019deep}
Oleg Ovcharenko, Vladimir Kazei, Mahesh Kalita, Daniel Peter, and Tariq Alkhalifah.
\newblock Deep learning for low-frequency extrapolation from multioffset seismic data.
\newblock {\em Geophysics}, 84(6):R989--R1001, 2019.

\bibitem{fang2020data}
Jinwei Fang, Hui Zhou, Yunyue Elita~Li, Qingchen Zhang, Lingqian Wang, Pengyuan Sun, and Jianlei Zhang.
\newblock Data-driven low-frequency signal recovery using deep-learning predictions in full-waveform inversion.
\newblock {\em Geophysics}, 85(6):A37--A43, 2020.

\bibitem{jin2021efficient}
Yuchen Jin, Wenyi Hu, Shirui Wang, Yuan Zi, Xuqing Wu, and Jiefu Chen.
\newblock Efficient progressive transfer learning for full-waveform inversion with extrapolated low-frequency reflection seismic data.
\newblock {\em IEEE Transactions on Geoscience and Remote Sensing}, 60:1--10, 2021.

\bibitem{luo2023low}
Renyu Luo, Jinghuai Gao, and Chuangji Meng.
\newblock Low-frequency prediction based on multiscale and cross-scale deep networks in full-waveform inversion.
\newblock {\em IEEE Transactions on Geoscience and Remote Sensing}, 61:1--11, 2023.

\bibitem{zhang2019regularized}
Zhen-Dong Zhang and Tariq Alkhalifah.
\newblock Regularized elastic full-waveform inversion using deep learning.
\newblock {\em GEOPHYSICS}, 84:R741--R751, 2019.

\bibitem{li2023self}
Yuanyuan Li, Tariq Alkhalifah, Jianping Huang, and Zhenchun Li.
\newblock Self-supervised pre-training vision transformer with masked autoencoders for building subsurface model.
\newblock {\em IEEE Transactions on Geoscience and Remote Sensing}, 2023.

\bibitem{wang2023prior}
Fu~Wang, Xinquan Huang, and Tariq~A. Alkhalifah.
\newblock A prior regularized full waveform inversion using generative diffusion models.
\newblock {\em IEEE Transactions on Geoscience and Remote Sensing}, 61:1--11, 2023.

\bibitem{wu2019inversionnet}
Yue Wu and Youzuo Lin.
\newblock Inversionnet: An efficient and accurate data-driven full waveform inversion.
\newblock {\em IEEE Transactions on Computational Imaging}, 6:419--433, 2019.

\bibitem{zhang2020data}
Zhongping Zhang and Youzuo Lin.
\newblock Data-driven seismic waveform inversion: A study on the robustness and generalization.
\newblock {\em IEEE Transactions on Geoscience and Remote sensing}, 58(10):6900--6913, 2020.

\bibitem{zeng2021inversionnet3d}
Qili Zeng, Shihang Feng, Brendt Wohlberg, and Youzuo Lin.
\newblock Inversionnet3d: Efficient and scalable learning for 3-d full-waveform inversion.
\newblock {\em IEEE Transactions on Geoscience and Remote Sensing}, 60:1--16, 2021.

\bibitem{wang2020velocity}
Wenlong Wang and Jianwei Ma.
\newblock Velocity model building in a crosswell acquisition geometry with image-trained artificial neural networks.
\newblock {\em Geophysics}, 85(2):U31--U46, 2020.

\bibitem{feng2021multiscale}
Shihang Feng, Youzuo Lin, and Brendt Wohlberg.
\newblock Multiscale data-driven seismic full-waveform inversion with field data study.
\newblock {\em IEEE Transactions on Geoscience and Remote Sensing}, 60:1--14, 2021.

\bibitem{ho2020denoising}
Jonathan Ho, Ajay Jain, and Pieter Abbeel.
\newblock Denoising diffusion probabilistic models.
\newblock {\em Advances in neural information processing systems}, 33:6840--6851, 2020.

\bibitem{dhariwal2021diffusion}
Prafulla Dhariwal and Alexander Nichol.
\newblock Diffusion models beat gans on image synthesis.
\newblock {\em Advances in neural information processing systems}, 34:8780--8794, 2021.

\bibitem{ho2022video}
Jonathan Ho, Tim Salimans, Alexey Gritsenko, William Chan, Mohammad Norouzi, and David~J Fleet.
\newblock Video diffusion models.
\newblock {\em Advances in Neural Information Processing Systems}, 35:8633--8646, 2022.

\bibitem{rombach2022high}
Robin Rombach, Andreas Blattmann, Dominik Lorenz, Patrick Esser, and Bj{\"o}rn Ommer.
\newblock High-resolution image synthesis with latent diffusion models.
\newblock In {\em Proceedings of the IEEE/CVF conference on computer vision and pattern recognition}, pages 10684--10695, 2022.

\bibitem{wang2024reconstructing}
Xinlei Wang, Zhiguo Wang, Zhe Xiong, Yang Yang, Chaobo Zhu, and Jinghuai Gao.
\newblock Reconstructing regularly missing seismic traces with a classifier-guided diffusion model.
\newblock {\em IEEE Transactions on Geoscience and Remote Sensing}, 62:1--14, 2024.

\bibitem{deng2024seismic}
Fei Deng, Shuang Wang, Xuben Wang, and Peng Fang.
\newblock Seismic data reconstruction based on conditional constraint diffusion model.
\newblock {\em IEEE Geoscience and Remote Sensing Letters}, 2024.

\bibitem{liu2024generative}
Qi~Liu and Jianwei Ma.
\newblock Generative interpolation via a diffusion probabilistic model.
\newblock {\em Geophysics}, 89(1):V65--V85, 2024.

\bibitem{zhu2023diffusion}
Donglin Zhu, Lei Fu, Vladimir Kazei, and Weichang Li.
\newblock Diffusion model for das-vsp data denoising.
\newblock {\em Sensors}, 23(20):8619, 2023.

\bibitem{zhang2024conditional}
Hao Zhang, Yuanyuan Li, and Jianping Huang.
\newblock Conditional denoising diffusion probabilistic model for seismic diffraction separation and imaging.
\newblock {\em IEEE Transactions on Geoscience and Remote Sensing}, 62:1--13, 2024.

\bibitem{li2024conditional}
Yuanyuan Li, Hao Zhang, Jianping Huang, and Zhenchun Li.
\newblock Conditional denoising diffusion probabilistic model for ground-roll attenuation.
\newblock {\em arXiv preprint arXiv:2403.18224}, 2024.

\bibitem{song2020denoising}
Jiaming Song, Chenlin Meng, and Stefano Ermon.
\newblock Denoising diffusion implicit models.
\newblock {\em arXiv preprint arXiv:2010.02502}, 2020.

\bibitem{choi2021ilvr}
Jooyoung Choi, Sungwon Kim, Yonghyun Jeong, Youngjune Gwon, and Sungroh Yoon.
\newblock Ilvr: Conditioning method for denoising diffusion probabilistic models.
\newblock {\em arXiv preprint arXiv:2108.02938}, 2021.

\bibitem{OpenFWIdeng2022}
Chengyuan Deng, Shihang Feng, Hanchen Wang, Xitong Zhang, Peng Jin, Yinan Feng, Qili Zeng, Yinpeng Chen, and Youzuo Lin.
\newblock Openfwi: Large-scale multi-structural benchmark datasets for full waveform inversion.
\newblock In S.~Koyejo, S.~Mohamed, A.~Agarwal, D.~Belgrave, K.~Cho, and A.~Oh, editors, {\em Advances in Neural Information Processing Systems}, volume~35, pages 6007--6020. Curran Associates, Inc., 2022.

\bibitem{nichol2021improved}
Alexander~Quinn Nichol and Prafulla Dhariwal.
\newblock Improved denoising diffusion probabilistic models.
\newblock In {\em International conference on machine learning}, pages 8162--8171. PMLR, 2021.

\end{thebibliography}

\end{document}